\documentclass[a4paper,english,twocolumn,nofootinbib,pre,superscriptaddress]{revtex4-1}

\usepackage{amsmath,amsfonts}
\usepackage{graphicx}
\usepackage{amssymb}
\usepackage{enumerate}
\usepackage{epsfig}
\usepackage{epstopdf}
\usepackage{mathrsfs}
\usepackage{dcolumn}
\usepackage{amsthm}
\usepackage[latin9]{inputenc}
\usepackage{textcomp}
\usepackage[colorlinks,citecolor=blue,linkcolor=blue]{hyperref}
\usepackage{babel}
\usepackage[all]{hypcap}

\newcommand{\E}{\mathcal{E}}

\newcommand{\ov}{\overline}

\newcommand{\ra}{\rangle}
\newcommand{\la}{\langle}

\newcommand{\be}{\begin{equation}}
\newcommand{\ee}{\end{equation}}

\newcommand{\N}{{\mathscr{N}}}

\newcommand{\tr}{{\rm Tr}}

\begin{document}

\title{Steady-state relations for a two-level system locally and relatively-strongly coupled to
 a generic many-body quantum chaotic  environment
}

\author{Hua Yan}
\affiliation{ Department of Modern Physics, University of Science and Technology of China,
 Hefei 230026, China}
\author{Jiaozi Wang}
\affiliation{ Department of Modern Physics, University of Science and Technology of China,
	Hefei 230026, China}
\affiliation{CAS Key Laboratory of Microscale Magnetic Resonance, 
University of Science and Technology of China, Hefei 230026, China}

\author{Wen-ge Wang}
\email{wgwang@ustc.edu.cn}
\affiliation{ Department of Modern Physics, University of Science and Technology of China,
	Hefei 230026, China}
\affiliation{CAS Key Laboratory of Microscale Magnetic Resonance, 
University of Science and Technology of China, Hefei 230026, China}

\date{\today}

\begin{abstract}
We study the long-time average of the reduced density matrix (RDM) of a two-level system as the central system,
which is locally coupled to a generic many-body  quantum chaotic system as the environment, 
under an overall Schr\"{o}dinger evolution. The system-environment interaction has a generic form with dissipation.
It is shown that, in addition to the exact relations due to unit trace and hermiticity,
an approximate relation exists among elements of the averaged RDM computed in the 
eigenbasis of the central system's Hamiltonian in some interaction regimes.
In particular, an explicit expression of the relation is derived for relatively strong interactions,
whose strength is above the mean level spacing of the environment, meanwhile,
remains small compared with the central system's level spacing.
Numerical simulations performed in a model with the environment as a defect Ising chain
confirm the analytical predictions.
\end{abstract}

\maketitle

\section{Introduction}\label{sect-intro}

\subsection{Motivation}

 Properties of small open quantum systems, which are coupled to large quantum environments,
 have attracted lots of attention and been studied extensively in recent decades in various fields of physics
 \cite{leggett1987dynamics,breuer2002, alicki2007,breuer2016,vega2017}.
 Such a system is described by its reduced density matrix (RDM)
 and may approach a steady state in many cases.
 For example,  it is now well known that  the phenomenon of decoherence,
 due to interaction with huge quantum environments,
 may happen in such a way that the RDM becomes approximately diagonal
 in a so-called preferred (pointer) basis
 \cite{Zurek81,zurek2003decoherence,Schloss05,JZKGKS03,wiseman2010quantum,PZ99,DK00}.
 For pure-dephasing interactions, decoherence has been studied well,
 with preferred bases given by the eigenbases of self-Hamiltonians
 \cite{PZ99,gorin2004connection,lidar2015,weiss2012}.
 In an opposite case with a negligible self-Hamiltonian and a strong interaction Hamiltonian,
 and with the environment undergoing a complex motion,
 it is the eigenbasis of the interaction Hamiltonian that gives the preferred basis
 \cite{Zurek81,braun2001universality,gorin2004connection}.

 The situation is much more complicated for generic dissipative system-environment interactions,
 due to the interplay of decoherence and relaxation.
 When the interaction is sufficiently weak such that the interplay can be studied by a first-order
 perturbation theory, it was found that the system's eigenbasis is approximately a preferred basis
 under a quantum chaotic environment \cite{wang2008entanglement}.
 Later, it was shown that  a preferred basis, if existing, should be given by
 the eigenbasis of a renormalized Hamiltonian,
 when the total system's eigenfunctions possess certain type of randomness \cite{pre14-ps}.
 This is in agreement with a generic expectation for Markovian processes described by Lindblad master
 equations, as exemplified in solvable models
\cite{breuer2002, alicki2007,lidar2015}.

 In contrast, as observed more recently, RDMs in the energy basis may possess
 nonnegligible offdiagonal elements at long times in various spin-boson models,
 when nonMarkovian effects due to the dynamics of the total system are taken into account
 \cite{GongJB12,addis2014coherence,roszak2015decoherence,zhang2015role,cakmak2017, guarnieri2018steady}.
 In these models, the bosons (as the environment), though coupled to the spin, do not interact with each other
 and, as a result,   do not undergo  a chaotic motion.
 In fact, a spin-boson model simulates the interaction of a two-level system (TLS)
 with an environment like a radiation field or phononic modes in a solid.

 There are many realistic situations that lie outside the scope of spin-boson models,
 e.g., a TLS that is locally coupled to some Ising chain with defects.
 Generically, one may consider a central system,
 which is locally coupled to an environment that undergoes a complex, even chaotic motion.
 An interesting question is whether a phenomenon similar to that discussed above may exist
 in such a situation.

\subsection{The problem to be addressed and organization of the paper}

 The exact problem to be addressed in this paper is as follows.
 Consider a TLS as the central system,
 which is coupled to a generic many-body quantum chaotic system as its environment.
 The interaction is of a generic type, except that it is local and not very strong.
 One question we ask is, besides the universal restrictions of unit trace and hermiticity of the RDM,
 whether any other generic relation/restriction may exist for a steady RDM of the TLS (possibly in an approximate way).
 Knowledge about such a property, if existing, should be useful in the study of many topics, e.g., 
 decoherence and quantum thermodynamics
 \cite{binder2019thermodynamics, goold2016role,wang2012statistical,wang2017decoherence}.

 We are to start from the overall Schr\"{o}dinger evolution of the total system
 and study the long-time average of the RDM.
 (Clearly, when a steady state of the TLS exists, its properties are given by the long-time averaged RDM.)
 A key point is to make use of the so-called eigenstate thermalization hypothesis (ETH)
 \cite{Deutch91,srednicki1994chaos,srednicki1999approach,rigol2008thermalization},
 which is usually expected to be valid for many-body quantum chaotic systems \cite{d2016quantum,Deutch18}.
 Based on this hypothesis, we'll show that, in an average sense,
 some dynamical properties of the interaction Hamiltonian can be related to the RDM.
 This makes it possible to derive some relations of the type discussed above.

 The paper is organized as follows.
 In Sec.~\ref{sect-setup}, we discuss basic properties of the systems to be studied,
 particularly of the environment as a many-body quantum chaotic system.
 Some preliminary and generic discussions are also given in this section,
 including a branch-structure of the time evolution of the total system
 and a formal expression for the long-time averaged RDM.
 In Sec.~\ref{sect-main}, we discuss the main result of this paper;
 we first consider a specific type of dissipative interaction and
 derive a relation among elements of the averaged RDM,
 then, we discuss a generic type of dissipative interaction with a relatively large strength.

 The averaged RDM under
 two other types of the interaction, namely, a very weak generic interaction and a pure-dephasing interaction,
 are discussed in Sec.~\ref{sect-derive-pd-vw}.
 In Sec.~\ref{sect-numerical}, numerical simulations are presented,
 to test the above-mentioned analytical predictions.
 Finally, conclusions and discussions are given in Sec.~\ref{sect-conclusion}.

\section{The set-up}\label{sect-setup}

 In this section, we discuss the basic framework within which our investigations will be carried out.
 Specifically, in Sec.~\ref{sect-Hamiltonian}, we discuss basic properties of and notations to be used for
 Hamiltonians and their eigenstates of the systems to be studied.
 In Sec.~\ref{sect-propt-chaos}, we discuss some  properties of the environment,
 which will be used in later discussions.
 Then, in Sec.~\ref{sect-branches}, we discuss a generic branch structure of
 the  Schr\"{o}dinger evolution of the total state.
 Finally, in Sec.~\ref{sect-longtime-RDM}, we derive a formal expression for the long-time averaged RDM,
 which will be used as a starting point in the derivation of the main result of this paper.

\subsection{Hamiltonians and their eigenstates}\label{sect-Hamiltonian}

 The TLS is denoted by $S$.
 Its environment, denoted by $\E$,
 is a generic many-body quantum chaotic system consisting of $N$ particles ($N\gg 1$).
 The Hilbert spaces of $S$ and $\E$ are denoted by $\mathcal{H}^S$ and $\mathcal{H}^\E$,
 respectively, with dimensions $d_S(=2)$ and $d_\E$.

 The Hamiltonian of the total system is written as
\begin{equation}
 H = H^S + H^I + H^\E,
\end{equation}
 where $H^S$ and $H^\E$ are the Hamiltonians of the two systems $S$ and $\E$, respectively,
 and $H^I$ represents an interaction Hamiltonian with a local feature.
 The interaction Hamiltonian is assumed to have a simple product form,
\begin{equation}\label{HI}
H^I = H^{IS}\otimes H^{I\E},
\end{equation}
 where $H^{IS}$ and $H^{I\E}$ are Hermitian operators acting on the two spaces
 $\mathcal{H}^S$ and $\mathcal{H}^\E$, respectively
 \footnote{At the end of Sec.~\ref{sect-conclusion}, we briefly discuss a possibility of generalizing
 results of this paper to a more generic form of the interaction Hamiltonian.}.

 For a local interaction, the environment $\E$ is further divided into a small part denoted by $\E_1$
 and a large part denoted by $\E_2$,
 such that the system $S$ is coupled to the small part $\E_1$ only.
 That is, $H^{I\E}$ has the following form,
\begin{gather}\label{HIE-local}
 H^{I\E} = H^{I\E_1} \otimes I^{\E_2},
\end{gather}
 where $H^{I\E_1}$ is an operator that acts on the Hilbert space of $\E_1$ and
 $I^{\E_2}$ indicates the identity operator on the Hilbert space of $\E_2$.

 Normalized eigenstates of $H^S$ are denoted by $|\alpha\rangle$, with energies $e^S_\alpha$
 ($\alpha =1,2$ and $e^S_2>e^S_1$), and those of $H^\E$ by $|i\rangle$ with energies $e_i$
 in the increasing-energy order,
\begin{eqnarray} \label{Sa}
H^S|\alpha\rangle &=& e^S_\alpha|\alpha\rangle, \\
H^\E|i\rangle &=& e_i|i\rangle, \label{Ei}
\end{eqnarray}
 where for brevity we have omitted a superscript $\E$ for $e_i$.
 Elements of the two operators $H^{IS}$ and $H^{I\E}$ in the bases given by the above states are written as
\begin{subequations}\label{HIS-alphabeta}
\begin{gather}
 H^{IS}_{\alpha \beta} = \la\alpha|H^{IS}|\beta\ra,
\\  H^{I\E}_{ij} = \la i | H^{I\E}|j\ra. \label{HIE-ij}
\end{gather}
\end{subequations}
 Eigenstates of the uncoupled Hamiltonian, denoted by $H^0$,
\begin{gather}\label{}
 H^0 = H^S + H^\E,
\end{gather}
 are written as $|\alpha\rangle |i\rangle$, in short, $|\alpha i \rangle$, satisfying
 $H^0|\alpha i\ra = E_{\alpha i}|\alpha i\ra$, where $E_{\alpha i} = e^S_\alpha + e  _i$.
 It is sometimes useful to write the basis states $|\alpha i\ra$ in the energy order,
 which we denote by $|E_r\ra$ with a one-to-one correspondence of $r\leftrightarrow (\alpha, i)$,
\begin{gather}\label{}
 H^0|E_r\ra = E_{r}|E_r\ra, \quad  E_r = E_{\alpha i}.
\end{gather}
 We use $\Delta_S$ to indicate the level spacing of the central system $S$,
\begin{gather}\label{}
 \Delta_S = e^S_2- e^S_1,
\end{gather}
 and use $s_\E$ to indicate the mean nearest-level spacing of the environment in the studied energy region.
 Due to the largeness of the environment, usually $\Delta_S \gg s_\E$, which is the case we consider here.

 Normalized eigenstates of the total system are denoted by $|n\ra$ with energies $E_{n}$
 in the increasing-energy order,
\begin{gather}\label{}
 H|n\rangle = E_{n}|n\rangle.
\end{gather}
 The expansions of $|n\ra$ in the bases of $|\alpha i\ra$ and $|E_r\ra$, with components
 denoted by $C_{\alpha i}^n$ and $C_r^n$, respectively, are written as
 \begin{equation}\label{|n>}
 |n\rangle = \sum_{\alpha i}C_{\alpha i}^n|\alpha i\rangle = \sum_{r}C_{r}^n|E_r\rangle.
\end{equation}
 These components give the EFs of $|n\ra$ in these two bases.

 The interaction strength can be characterized by a local Hilbert-Schmidt
 norm of $H^I$, denoted by $\| H^{I}_L \|$ with the subscript ``$L$'' standing for ``local'', i.e.,
\begin{gather}\label{norm-HI}
 \| H^{I}_L \| = \sqrt{\tr_S(H^{IS\dag} H^{IS}) \tr_{\E_1}(H^{I\E_1 \dag} H^{I\E_1})  }.
\end{gather}
 It proves convenient to introduce a running parameter $\lambda$ for the interaction strength
 and write the interaction Hamiltonian as $H^I(\lambda)$,
 with $ \| H^{I}_L \|$ increasing with $\lambda$.
 In our study of dissipative interactions, we divide the 
 whole interaction regime into four subregions: very weak, relatively weak, 
 relatively strong, and very strong.
\begin{itemize}
  \item In the \emph{very weak} regime, the EF of each state $|n\ra$ contains only one big component, 
  with $|C^n_r|^2 \simeq 1$.
  \item In the \emph{relatively weak} regime, the EF  $C^n_r$ of each state $|n\ra$ contains a few big components.
  \item In the \emph{relatively strong} regime, the EF  $C^n_r$ of each state $|n\ra$ contains many significant components,
  while, remains narrow in energy.
  \item In the \emph{very strong} regime, the EF  $C^n_r$ of each state $|n\ra$ contains many significant components
  and is wide in energy.
\end{itemize}
 Further explanation to the meaning of narrowness of EFs in our study will be given in Sec.~\ref{sect-main-content2}.
 Loosely speaking, $\frac{1}{d_S d_{\E_1}} \| H^{I}_L(\lambda)\| < s_\E $ in the very weak regime,
 where $d_{\E_1}$ indicates the dimension of the Hilbert space of $\E_1$,
 while, $\| H^{I}_L(\lambda) \| \gtrsim \Delta _S$ in the very strong regime. 
 \footnote{For a quantitative analysis in
 the scale of the offdiagonal elements of $H^I$ in the basis of $|E_r\ra$
 in a total system of the type studied here,  see the appendix A of  Ref.~\cite{pre12-sta}. }

\subsection{Some properties of the chaotic environment}\label{sect-propt-chaos}

 In this section, we discuss some properties to be used in later discussions,
 which are related to the quantum chaotic feature of the environment $\E$.
 Firstly, the most-often-used criterion for quantum chaos is given by the spectral statistics;
 that is, a system is said to be a quantum chaotic system, if
 its spectral statistics is in agreement with predictions of the random matrix theory (RMT)
 \cite{casati79,Bohigas84,berry85,Sieber-Richter00,Haake04}.
\footnote{For a system that has a conserved observable, which is independent of the Hamiltonian, this criterion is
 usable within each eigensubspace of the observable.}
 According to this definition, neither the environmental levels $e_i$ nor their spacings have any degeneracy
 (neglecting the very special case of accidental degeneracy).
 For a generic value of $\Delta_S$ and under a generic type of the system-environment interaction,
 the spectrum of the total system has this property, too.

 Secondly, although the exact condition under which  the ETH ansatz
 proposed in Ref.\cite{srednicki1999approach}  is applicable is still unclear,
 it is usually expected valid for local operators of many-body quantum chaotic systems
 \cite{d2016quantum,Deutch18}.
 Hence, we assume that this ansatz is applicable to the operator $H^{I\E}$.
 According to this hypothesis,
 (i) the diagonal element $H^{I\E}_{ii}$ on average varies slowly with the eigenenergy $e_i$;
 (ii) the fluctuation of $H^{I\E}_{ii}$ possesses certain random feature and is very small,
 scaling as $e^{-S(e)/2}$, where $S(e)$ is proportional to the particle number $N$ of $\E$
 and is related to the microcanonical entropy in a semiclassical treatment;
 and (iii) the offdiagonal elements $H^{I\E}_{ij}$ with $i\ne j$ behave in a way similar to the fluctuation of $H^{I\E}_{ii}$
 \cite{Deutch91,srednicki1994chaos,rigol2008thermalization,srednicki1999approach,
 d2016quantum, garrison2018does}.
 These predictions are written in the following concise form, referred to as the ETH ansatz,
 \begin{equation}\label{ETH}
 H^{I\E}_{ij} = {h}(e_i) \delta_{ij} + e^{-S(e_i)/2}g(e_i,e_j)R_{ij},
\end{equation}
 where  ${h}(e)$ is a slowly-varying function of $e$,
 $g(e_i,e_j)$ is some smooth function of its variables ($|g|$ being not large),
 and the quantity $R_{ij}$ has certain random feature with
 a normal distribution (zero mean and unit variance).

 Thirdly, we discuss an estimate to widths of main-body regions of the EFs of $|n\ra$ in the basis of $|E_r\ra$.
 Here, the main-body region of an EF $C^n_r$ refers to a connected region of the energy $E_r$,
 which contains the main population of the EF, in other words, over which the sum of $|C^n_r|^2$
 is close to $1$. 
 According to a result of Ref.\cite{pre12-sta} (appendix B there),
 the maximum width of the main-body regions, denoted by $w_{\rm mb}$, scales as
 $w_{\rm mb} \propto \frac{1}{\Delta_{\cal E}}$,
 where $\Delta_\E$ represents the total energy scale of the environment $\E$
 and the the subscript ``mb'' stands for ``main body''.
 \footnote{ This result of Ref.\cite{pre12-sta} was obtained
 by making use of a first-order perturbation-theory treatment to the long tails of EFs.
 There, it was pointed out that validity of  this treatment is based on a
 generalized Brillouin-Wigner perturbation theory \cite{WIC98,pre00-02-GBW}.
 A detailed justification of the validity will be given in a different paper \cite{EF-semp}.}
 The value of $\Delta_{\E}$ usually has a power-law dependence on $N$ and, hence,
 is large for a large environment, implying narrowness of the EFs.

\subsection{A branch-structure of the total evolution}\label{sect-branches}

 We consider an initial product state of the total system at a time $t=0$, written as
\begin{equation}
 |\Psi(0)\rangle = |\phi_S\rangle \otimes |\E_0\rangle.
 \end{equation}
 Here, $|\phi_S\rangle$ indicates an arbitrary normalized state of the central system $S$,
\begin{gather}\label{phiS}
 |\phi_S\ra = \sum_\alpha c_{0\alpha} |\alpha\rangle,
\end{gather}
 and $|\E_0\ra$ is a typical state of the environment, which lies within a narrow energy shell
 denoted by $\Gamma^{\cal E}_0$.
 The shell is centered at an energy $e_0$, with a width
 $\delta e_0$, namely, $\Gamma^{\cal E}_0 = [e_0-\delta e_0/2,e_0+\delta e_0/2 ]$.
\footnote{ The energy shell is not close to edges of the spectrum. }
 Expanded in the basis $|e_i\ra$, $|\E_0\ra$ is written as
\begin{equation}\label{E0}
|\E_0\rangle = \N_0 \sum_{e_i\in \Gamma^{\cal E}_0} c_{0i}|i\rangle,
\end{equation}
 where the real and imaginary parts of $c_{0i}$ are independent Gaussian random
 variables, with mean zero and variance $1/2$, and $\N_0$ is the normalization factor.
 We use $N_{\Gamma_0^\E}$ to indicate the number of levels within the shell $\Gamma_0^\E$.
 Although being narrow, $\Gamma_0^\E$ is assumed to contain a large number of energy levels.
 It is seen that $\N_0^2 \simeq 1/N_{\Gamma_0^\E}$.

 The time evolution of the total system is given by 
\begin{gather}\label{Psi-t}
 |\Psi(t)\rangle  = e^{-iHt/\hbar }|\Psi(0)\rangle.
\end{gather}
 Writing $|\Psi(t)\ra$ in the following expansion,
\begin{gather}\label{Psi-expan}
|\Psi(t)\rangle = \sum_{\alpha=1}^{d_S} |\alpha\rangle|\E_\alpha(t)\rangle,
\end{gather}
 we call $|\E_\alpha(t)\rangle$ the environmental \emph{branches}  of $|\Psi(t)\ra$.
 These branches, given by
\begin{gather}\label{}
 |\E_\alpha(t)\rangle = \la \alpha|\Psi(t)\rangle,
\end{gather}
 are usually not normalized.
 From the Schr\"{o}dinger evolution of the total system  in Eq.(\ref{Psi-t}), one finds that
\begin{equation} \label{alpha-t}
 i\hbar \frac{d|\E_\alpha(t)\rangle}{dt} = H_{\E\alpha}^{\rm eff}|\E_\alpha(t)\rangle
 + \sum_{\beta \ne \alpha} H^{I}_{\alpha\beta}|\E_\beta(t)\rangle,
\end{equation}
 where $H^I_{\alpha\beta}$ and $H_{\E\alpha}^{\rm eff}$ are operators that act on the Hilbert space
 of the environment, as defined below,
\begin{gather}\label{}
   H^I_{\alpha\beta} := H^{IS}_{\alpha\beta} H^{I\E},
 \\ H_{\E\alpha}^{\rm eff} := H^\E +  H^I_{\alpha\alpha} + e^S_\alpha.
\end{gather}

 By definition, the RDM of the system $S$, denoted by $\rho^S(t)$,
 is written as $\rho^S(t) = \text{Tr}_\E \rho(t)$, where $\rho(t) = |\Psi(t)\rangle\langle\Psi(t)|$.
 It is easy to check that elements of the RDM in the basis $\{|\alpha\ra \}$, namely
 $\rho_{\alpha\beta}^S(t) \equiv \langle \alpha|\rho^S(t)|\beta\rangle$,
 have the following expression,
\begin{equation}\label{rho-abt}
\rho_{\alpha\beta}^S(t) = \langle \E_\beta(t)|\E_\alpha(t)\rangle.
\end{equation}
 Let us expand the environmental branches $|\E_\alpha(t)\rangle$ on the basis of $|i\rangle$, obtaining
\begin{gather}\label{Ealpha(t)}
|\E_\alpha(t)\rangle = \sum_i f_{\alpha i }(t)|i\rangle.
\end{gather}
 Then, elements of the RDM have the following expression,
\begin{equation}\label{rho-abt-CC}
 \rho_{\alpha\beta}^S(t) = \sum_i f^*_{\beta i}(t)f_{\alpha i}(t).
\end{equation}
 The expansion coefficients, $f_{\alpha i }(t) = \la i|\E_\alpha(t)\rangle = \la \alpha i|e^{-iHt/\hbar }|\Psi(0)\rangle$,
 have the following explicit expression,
\begin{gather}\label{fat-evol}
 f_{\alpha i }(t)  =  \N_0  \sum_{n,\alpha'} \sum_{e_j\in\Gamma^{\cal E}_0}  c_{0\alpha'} c_{0j}
 e^{-iE_{n}t/\hbar }  C_{\alpha i}^n  C_{\alpha' j}^{n*}.
\end{gather}

 It proves useful to further divide each branch $|\E_\alpha(t)\ra$ into subbranches,
 according to the initial components they come from.
 Such a subbranch is denoted by $|\E^{(\alpha')}_\alpha(t)\ra$,
 with the label  for the initial component denoted by $\alpha'$
 and written within a pair of parenthesis in the upper position.
 Specifically,
 \begin{gather}\label{Eabt}
 |\E_\alpha(t)\ra = \sum_{\alpha'} |\E^{(\alpha')}_\alpha(t)\rangle,
 \end{gather}
 where
 \begin{gather}\label{Ebeta-t}
 |\E^{(\alpha')}_\alpha(t)\rangle = c_{0\alpha'}  \la \alpha|e^{-iHt/\hbar} |\alpha'\rangle |\E_0\ra.
 \end{gather}
 The total time evolution is then written as
 \begin{gather}\label{}
 |\Psi(t)\rangle = \sum_{\alpha,\alpha'} |\alpha\rangle |\E^{(\alpha')}_\alpha(t)\rangle.
 \end{gather}
 Expanding $|\E^{(\alpha' )}_\alpha(t)\ra$ in the basis states $|i\ra$, one writes
 \begin{gather}\label{Ealpha-b(t)}
 |\E^{(\alpha')}_\alpha(t)\rangle = \sum_i f^{(\alpha')}_{\alpha i }(t)|i\rangle,
 \end{gather}
 where
\begin{gather}\label{fba-t}
 f^{(\alpha')}_{\alpha i }(t) = \la i |\E^{(\alpha')}_\alpha(t)\rangle
 = c_{0\alpha'}  \la \alpha i|e^{-iHt/\hbar} |\alpha'\rangle |\E_0\ra.
\end{gather}
 It is easy to see that
\begin{gather}
 f^{(\alpha')}_{\alpha i }(t) = \N_0 c_{0\alpha'} \sum_n \sum_{e_j\in\Gamma^{\cal E}_0} c_{0j}
 e^{-iE_{n}t/\hbar }  C_{\alpha i}^n  C_{\alpha' j}^{n*}.\label{fba-t-sum}
\end{gather}

\subsection{A formal expression for the long-time averaged RDM}\label{sect-longtime-RDM}

 Making use of Eq.(\ref{alpha-t}), it is straightforward to find that
 the elements $\rho_{\alpha\beta}^S(t)$ satisfy the following equation,
\begin{eqnarray}\label{drhot}
-i\hbar \frac{d\rho_{\alpha\beta}^S(t)}{dt} &=& W_{\alpha\beta}^{(1)} + W_{\alpha\beta}^{(2)},
\end{eqnarray}
 where
\begin{subequations}\label{W1-2}
\begin{eqnarray}\label{w1}
 W_{\alpha\beta}^{(1)} &=& (e_\beta^S-e_\alpha^S)\rho_{\alpha\beta}^S(t),\\
 \label{w2}
 W_{\alpha\beta}^{(2)} &=& \sum_{\gamma=1}^{d_S} \big(H^{IS}_{\gamma\beta}
 F_{\gamma\alpha}(t) - H_{\alpha \gamma}^{IS} F_{\beta \gamma}(t) \big).
\end{eqnarray}
\end{subequations}
 Here, $F_{\alpha\beta}(t)$ indicate the quantities defined below,
\begin{equation}\label{HIE-ab}
F_{\alpha\beta}(t) := \langle \E_\alpha(t)|H^{I\E}|\E_\beta(t)\rangle.
\end{equation}
 In terms of subbranches, $F_{\alpha\beta}(t)$  is written as
 \begin{gather}\label{Fab-aba'b'}
 F_{\alpha\beta}(t) =  \sum_{\alpha' \beta'}  F_{\alpha\beta}^{(\alpha' \beta')}(t),
 \end{gather}
 where
 \begin{gather}\label{Faba'b'}
 F_{\alpha\beta}^{(\alpha' \beta')}(t)
 :=  \langle \E_\alpha^{(\alpha')}(t)|H^{I\E}|\E_\beta^{(\beta')}(t)\rangle.
 \end{gather}

 We use an overline to indicate the long-time average of a term.
 For example, the long-time average of the RDM is written as $\overline{\rho}^S$,
\begin{equation}\label{ov-rho}
\overline{\rho}^S   =\lim_{t\to\infty} \frac{1}{t}\int_0^t \rho^S(t')dt'.
\end{equation}
 Since the elements $\rho_{\alpha\beta}^S(t)$ have bounded values,
 the long-time average of $d\rho_{\alpha\beta}^S(t)/{dt}$ must be zero,
 $\ov{{d\rho_{\alpha\beta}^S(t)}/{dt}}=0$.
 Then, Eq.(\ref{drhot}) gives that
\begin{gather}\label{ov-WW=0}
 \ov {W}_{\alpha\beta}^{(1)} + \ov {W}_{\alpha\beta}^{(2)} =0.
\end{gather}
Substituting the explicit expressions of ${W}_{\alpha\beta}^{(1)}$
 and ${W}_{\alpha\beta}^{(2)}$ in Eq.(\ref{W1-2}) into Eq.(\ref{ov-WW=0}), one finds that
\begin{equation}\label{rho-S-0}
\overline{\rho}_{\alpha\beta}^S = \frac{1}{e_\beta^S-e_\alpha^S}
\sum_{\gamma=1}^{d_S} \Big[H_{\alpha \gamma}^{IS} \ov{ F}_{\beta \gamma}-H^{IS}_{\gamma\beta}
 \ov{ F}_{\gamma\alpha} \Big] \ (\alpha \ne \beta).
\end{equation}
 It is easy to verify that
\begin{align}
\label{hie-aa-1}
\ov{F}_{\alpha\beta} &= \sum_{i, j}\ov{f_{\alpha i}^*f_{\beta j}} H^{I\E}_{ij}, \quad \forall \alpha, \beta.
\end{align}

 For a TLS, Eq.(\ref{rho-S-0}) has a simpler form, i.e.,
\begin{gather}\label{tls-rho}
\overline{\rho}_{\alpha\beta}^S  = \eta_d \ov{F}_{\beta \alpha}+\eta_r  (\ov{F}_{\beta\beta}- \ov{F}_{\alpha\alpha})
 \ \quad \  (\alpha \ne \beta),
\end{gather}
 where
\begin{gather}\label{eta-dr}
\eta_{d} :=\frac{H_{\alpha\alpha}^{IS}-H_{\beta\beta}^{IS}}{e_{\beta}^{S}-e_{\alpha}^{S}},
\quad\eta_{r} :=\frac{H_{\alpha\beta}^{IS}}{e_{\beta}^{S}-e_{\alpha}^{S}}.
\end{gather}
 The quantity $\eta_d$ gives a relative measure for the strength of dephasing,
 while, $\eta_r$ gives a relative measure for the strength of relaxation or dissipation.
 (The subscript ``$d$'' of $\eta_d$ stands for ``dephasing'' and ``$r$'' of $\eta_r$ for ``relaxation''.)
 For a given operator $H^{IS}$, $\eta_d$ and $\eta_r$  can be regarded as given parameters;

 Equation (\ref{tls-rho}) will be used as the starting point in the next section,
 in the derivation of the main result of this paper. 
 This equation suggests that one may focus on the quantities $\ov{F}_{\beta \alpha}$.
 In fact, we are to show that, under conditions to be specified,  the quantities $\ov{F}_{\beta \alpha}$
 can be related to  elements of $\ov \rho^S$.

\section{The main result }\label{sect-main}

 The main result of this paper consists of two parts.
 The first part is derived in Sec.~\ref{sect-constant-h}, which gives a relation among elements of $\ov\rho^S$
 in the specific case that $h(e)$ is a constant function.
 The second part is for nonconstant $h(e)$.
 Its contents are given in Sec.~\ref{sect-main-content2}, with some preliminary discussions.
 In Sec.~\ref{sect-Faa}, we derive a relation between $\ov F_{\alpha \alpha}$ and diagonal elements of $\ov\rho^S$,
 then, making use of this relation, in Sec.~\ref{sect-derive-main} we give a proof for the second part of the main result.

\subsection{The first part of the main result}\label{sect-constant-h}

 To study the quantities $\ov{F}_{\alpha\beta}$,
 one may substitute Eq.(\ref{ETH}) into Eq.(\ref{hie-aa-1}) and get that
\begin{gather}\label{Fab-pd}
 \ov{F}_{\alpha\beta} = \sum_{i}\ov{f_{\alpha i}^*f_{\beta i}} h(e_i) + \Delta_{\alpha\beta},
\end{gather}
where
\begin{align}
 \Delta_{\alpha\beta} & = \sum_{i\neq j} \ov{f_{\alpha i}^*f_{\beta j}} e^{-S(e)/2}g(e_i,e_j)R_{ij}.\label{Delta-ab}
\end{align}
 Noting the normalization condition of $\sum_{\alpha, i} |f_{\alpha i}|^2 =1$
 and the randomness of $R_{ij}$, one finds that $|\Delta_{\alpha\beta}|$ usually have the same order
 of magnitude as $\la |g| \ra e^{-S(e)/2}$, i.e.,
\begin{gather}\label{D-ab-gS}
 |\Delta_{\alpha\beta}| \sim  \la |g| \ra e^{-S(e)/2},
\end{gather}
 where $\la |g| \ra$ indicates the average value of $|g(e_i,e_j)|$.
 Since $S(e) \propto N$ and $\la |g| \ra$ is not large,
 $\Delta_{\alpha \beta}$ is a small quantity for a large particle number $N$.

 Now, we discuss conditions under which the terms $\Delta_{\alpha \beta }$ can be neglected,
 when Eq.(\ref{Fab-pd}) is used in the derivation of relations among elements of $\ov\rho$ from Eq.(\ref{tls-rho}).
 To this end, we note the following estimate,
 which can be obtained from Eqs.(\ref{ETH}) and (\ref{D-ab-gS}), i.e.,
\begin{gather}\label{HIE-fluc-Delta}
 \la H^{I\E}_{\rm fluc} \ra \sim |\Delta_{\alpha\beta}|,
\end{gather}
 where $\la H^{I\E}_{\rm fluc} \ra$ indicates the average value of $|H^{I\E}_{ij}|$
 for the second item on the right-hand side (rhs) of Eqs.(\ref{ETH}), i.e.,
 for the fluctuation part of $H^{I\E}_{ii}$ and for the offdiagonal elements $H^{I\E}_{ij}$.

 Let us first consider the case of $\beta = \alpha$.
 Since $h(e)$ is a slowly-varying function and the energy region of relevance  is not wide,
 the value of $\sum_{i}\ov{f_{\alpha i}^*f_{\alpha i}} h(e_i)$ is close to $\la h \ra \ov\rho^S_{\alpha \alpha}$
 [cf.~Eq.(\ref{rho-abt-CC})],
 where $\la h \ra$ indicates the average value of $h(e_i)$ within the environmental energy region concerned.
 Making use of this estimate and Eq.(\ref{Fab-pd}), one finds that
\begin{gather}\label{F-F-hrhod}
 \ov{F}_{\beta\beta}- \ov{F}_{\alpha\alpha}
 \simeq  \la h \ra  (\ov\rho^S_{\beta\beta}- \ov\rho^S_{\alpha\alpha}) - (\Delta_{\beta\beta} -\Delta_{\alpha\alpha}).
\end{gather}
 Furthermore, due to the term $e^{-S}$ in $\la H^{I\E}_{\rm fluc} \ra$,
 when $|\la h \ra|$ is not very small and $N$ is sufficiently large, one has
\begin{gather}\label{hc>>HIE}
 |\la h \ra| \gg \la H^{I\E}_{\rm fluc} \ra.
\end{gather}
 When the temperature is not very high,
 the value of $|\ov\rho^S_{\beta\beta}- \ov\rho^S_{\alpha\alpha}|$ is not small.
 Then, making use of Eqs.(\ref{HIE-fluc-Delta}) and (\ref{hc>>HIE}), one finds that
 $|\la h \ra  (\ov\rho^S_{\beta\beta}- \ov\rho^S_{\alpha\alpha})| \gg |\Delta_{\beta\beta} -\Delta_{\alpha\alpha}|$.
 Thus, from Eq.(\ref{F-F-hrhod}) it is seen that,
 as far as Eq.(\ref{tls-rho}) is used, the two items $\Delta_{\alpha \alpha}$ of $\alpha =0,1$
 for $\ov{F}_{\alpha\alpha}$ can be neglected.
 \footnote{In the derivation of Eq.(\ref{F-F-hrhod}), $h(e_i)$ are taken as
 $\la h \ra$ for the energy regions concerned.
 This approximation is used here 
 for the purpose of showing that  $\Delta_{\alpha \alpha}$
 can be neglected under the condition of (\ref{hc>>HIE}),
 and it is not used in the derivation of the second part of the main result. }

 Next, we consider the case of $\alpha \ne \beta$.
 When the temperature is not very high and the value of $|\eta_r|$ is not much smaller than $|\eta_d|$,
 making use of Eq.(\ref{hc>>HIE}), one gets that
\begin{gather}\label{hrho>>dHIE}
 |\eta_r  \la h \ra  (\ov\rho^S_{\beta\beta}- \ov\rho^S_{\alpha\alpha})| \gg |\eta_d| \la H^{I\E}_{\rm fluc} \ra .
\end{gather}
 This implies that
 $|\eta_r  (\ov{F}_{\beta\beta}- \ov{F}_{\alpha\alpha})| \gg  |\eta_d \Delta_{\beta \alpha}|$
 and, as a result, $\Delta_{\beta\alpha}$ for the term of $\ov{F}_{\beta\alpha}$ in Eq.(\ref{tls-rho})
 can be neglected.

 Summarizing the above discussions,  when  the temperature is not very high,
 under the conditions of Eqs.(\ref{hc>>HIE})-(\ref{hrho>>dHIE}),
 the quantities $ \ov{F}_{\alpha \beta}$ can be approximately written as
\begin{equation}\label{ovFab-app-1}
 \ov{F}_{\alpha \beta} \simeq \sum_{i}\ov{f_{\alpha i}^*f_{\beta i}} h(e_i), \quad \forall \alpha, \beta,
\end{equation}
 as far as being used in Eq.(\ref{tls-rho}).
 We would emphasize again: Eq.(\ref{hc>>HIE}) is valid, if $|\la h \ra|$ is not very small and $N$ is sufficiently large;
 meanwhile, Eq.(\ref{hc>>HIE}) implies Eq.(\ref{hrho>>dHIE}), when $|\eta_r|$ is not much smaller than $|\eta_d|$
 (i.e., the dissipation effect is not very weak compared with the dephasing effect).

 Finally, we discuss the special case that $h(e)$ is a constant function, with a value denoted by $h_c$.
 From Eqs.(\ref{ovFab-app-1}) and (\ref{rho-abt-CC}), one finds that the quantity
 $\ov{F}_{\alpha\beta}$ is related to the RDM element $\ov \rho_{\beta\alpha}^S$ by
\begin{gather}\label{Fab-rho-special}
 \ov{F}_{ \alpha\beta} \simeq h_c \ov \rho_{\beta\alpha}^S, \quad \forall \alpha, \beta.
\end{gather}
 Substituting Eq.(\ref{Fab-rho-special}) into Eq.(\ref{tls-rho}), one finds that
 the elements of $\ov\rho^S$ satisfy the following relation,
\begin{gather}\label{ov-rho-gen-HIS-special}
 \overline{\rho}_{\alpha\beta}^S \simeq \frac{ \eta_r h_c}{1-\eta_d h_c}
 (\ov{\rho}_{\beta\beta}^S-\ov{\rho}_{\alpha\alpha}^S) \quad (\alpha \ne \beta).
\end{gather}
 The relation in Eq.(\ref{ov-rho-gen-HIS-special}) is the first part of the main result of this paper.

\subsection{Contents of the second part of the main result}\label{sect-main-content2}

 When $h(e)$ is not a constant, the situation is much more complicated than that discussed above.
 The point lies in whether an approximate relationship may exist between the rhs of Eq.(\ref{ovFab-app-1}) and
 elements of $\ov \rho^S$.
 We are to show that such a relation may indeed exist, as the second part of the main result of this paper.
 In this section, we describe contents of this second part.
 Its derivation will be given in the two sections following this one.

 The main content is that, under the prerequisites to be stated below, the following relation holds,
\begin{gather}\label{main-result}
 \overline{\rho}_{\alpha\beta}^S \simeq \frac{ \eta_r}{1-\eta_d h_0}
 (h_\beta\ov{\rho}_{\beta\beta}^S-h_\alpha\ov{\rho}_{\alpha\alpha}^S) \quad (\beta \ne \alpha),
\end{gather}
 where
\begin{gather}\label{}
 h_0 = h(e_0), \qquad  h_\alpha =  h(e_\alpha),
 \\ e_\alpha = e_0 + |c_{0\beta}|^2(e^S_\beta-e^S_\alpha) \quad \text{$(\beta\ne \alpha)$}.
\end{gather}
 There are four major prerequisites, as listed below.
\begin{itemize}
   \item[P1.] Smallness of the initial width $\delta e_0$.
   \item[P2.] Narrowness of the EFs $C^n_r$.
   \item[P3.] For each $\alpha$, the fitting function $B_\alpha(E)$ for the following quantities,
\begin{gather}\label{Bn}
 B_{\alpha }(E_{n}) := \sum_{i} |C_{\alpha i}^n|^2 = \la n| (|\alpha\ra \la \alpha|)|n\ra ,
\end{gather}
  changes slowly with $E$.
  \item [P4.] Validity of Eqs.(\ref{hc>>HIE}) and (\ref{hrho>>dHIE}).
 \end{itemize}
 Some minor prerequisites will also be used in the derivation of Eq.(\ref{main-result}),
 such as that $\Delta_S$ is not large and $N$ is large (see Sec.~\ref{sect-derive-main}).

 Below are further explanations to the above-discussed major prerequisites.
\\ (i) Due to the slow-variation feature of the function $h(e)$,
the smallness of $\delta e_0$ required in P1 implies that
 $h(e)$ may be treated as a constant within a width $\delta e_0$.
 Note that we do not require such a smallness of $\Delta_S$; in other words,
 $h(e)$ does not need to be approximately a constant within a width $\Delta_S$.
\\ (ii) The narrowness of EFs stated in P2 means that the value of $w_{\rm mb}$,
 which is defined in the last paragraph of Sec.~\ref{sect-propt-chaos},  is much smaller than
 all other energy scales considered in our discussions, e.g., $w_{\rm mb} \ll \delta e_0$.
 We take this as a prerequisite (P2), not as a prediction of Ref.\cite{pre12-sta}, for two reasons:
 One is that the exact conditions given there for the narrowness of EFs are quite complicated;
 the other is that the treatment given there is not complete, as explained in a footnote of Sec.~\ref{sect-propt-chaos}.
 \\ (iii) For  P2 to be satisfied, the interaction can not be very strong.
 According to analysis given in Ref.\cite{pre12-sta}
 \footnote{In Ref.\cite{pre12-sta}, EFs of the environment in an integrable basis is assumed to occupy 
 the whole energy region, as predicted by the RMT. 
 Making use of the ETH ansatz, analysis given there can be made simpler and
 be applied to realistic models with local interactions, whose EFs are so wide.  },
 for an environment of the type studied here,
 when the particle number $N$ is sufficiently large,
 the EFs $C^n_r$ are very narrow under the condition that $\| H^{I}_L(\lambda) \| \ll \Delta _S$. 
\\ (iv) For the condition P3 to be satisfied, each EF should contain many levels, i.e., $w_{\rm mb} \gg s_\E$.
 This requires that the interaction can not be weak.
 In fact, in the uncoupled case with $|n\ra = |\alpha i\ra$,
 $B_{\alpha }(E_{n})$ fluctuates between $0$ and $1$.
\\ (v) According to the above discussions, the two requirements of P2 and P3 are satisfied in the relatively strong regime
 of the interaction, when $N$ is sufficiently large
 \footnote{A full analysis of the scaling behavior of $w_{\rm mb} $ on $N$ needs further investigations.
 For a partial analysis, see   Ref.\cite{pre12-sta}.}
\footnote{ In this interaction regime, it is reasonable to expect that
 the total system should be also a quantum chaotic system under a generic form of the interaction Hamiltonian.
 Although this expectation is in agreement with experiences obtained from numerical simulations,
 a rigorous proof is still lacking.}.
 One should note that, since $\| H^{I}_L(\lambda) \| \ll \Delta _S$ is not a necessary condition for 
 the narrowness of the EFs $C^n_r$, the second part of the main result may in some cases be valid above (not quite above)
 the relatively strong regime of interaction. 
\\ (vi) Under P4, as discussed previously, Eq.(\ref{ovFab-app-1}) is applicable
 as far as being used in Eq.(\ref{tls-rho}).

\subsection{A relation between $\ov F_{\alpha\alpha}$ and $\ov \rho^S_{\beta\beta}$}\label{sect-Faa}

 In this section, we derive an approximate expression of $\ov F_{\alpha\alpha}$, given by
 diagonal elements of $\ov\rho^S$.
 We do this by four steps.

 Firstly, we show that $\ov F_{\alpha\alpha}$ can be approximately computed from
 $\overline{F}_{\alpha\alpha}^{(\alpha^{\prime}\alpha^{\prime})}$ of $\alpha' =1,2$.
 To this end, we substitute Eq.(\ref{Ealpha-b(t)}) into Eq.(\ref{Faba'b'}).
 Making use of the ETH ansatz in Eq.(\ref{ETH}), this gives an expression of $\ov F_{\alpha\beta}^{(\alpha' \beta')}$
 that has a form similar to  Eq.(\ref{Fab-pd}).
 Then, following arguments similar to those given in Sec.~\ref{sect-constant-h}, under the condition P4
 that guarantees the usage of Eq.(\ref{ovFab-app-1}),
 we find the following expression of $\overline{F}_{\alpha\beta}^{(\alpha^{\prime}\beta^{\prime})}$,
\be\label{Fab-sub}
 \overline{F}_{\alpha\beta}^{(\alpha^{\prime}\beta^{\prime})}\simeq
 \sum_{i}\overline{(f_{\alpha i}^{(\alpha^{\prime})})^{*}f_{\beta i}^{(\beta^{\prime})}}h(e_{i}).
\ee
 Since the EFs of $|n\ra$ contain many significant components (see discussions about P3),
 many of the terms $f_{\alpha i}^{(\alpha^{\prime})}$ and $f_{\beta i}^{(\beta^{\prime})}$ have significant values.
 Moreover, note that the phases of $f_{\alpha i}^{(\alpha^{\prime})}(t)$
 and $f_{\alpha i}^{(\beta^{\prime})}(t)$ with $\alpha' \ne \beta'$ usually do not completely match for long times,
 while, $\overline{(f_{\alpha i}^{(\alpha^{\prime})})^{*}f_{\alpha i}^{(\alpha^{\prime})}}
 = \overline{|f_{\alpha i}^{(\alpha^{\prime})}|^2}$.
 Then,  one gets that
\begin{gather}\label{}
 \left| \overline{F}_{\alpha\alpha}^{(12)}  + \overline{F}_{\alpha\alpha}^{(21)} \right|
 \ll  \overline{F}_{\alpha\alpha}^{(11)} + \overline{F}_{\alpha\alpha}^{(22)}.
\end{gather}
 Hence,
\begin{gather}\label{Faa-few}
 \overline{F}_{\alpha\alpha}\simeq  \sum_{\alpha'} \overline{F}_{\alpha\alpha}^{(\alpha^{\prime}\alpha^{\prime})}.
\end{gather}

 Secondly, we derive a simplified expression for $\overline{ \left| (f_{\alpha i}^{(\alpha')}) \right|^2}$.
 Making use of Eqs.(\ref{fba-t-sum}), one gets that
 \begin{gather}
 \ov{|f^{(\alpha')}_{\alpha i}(t)|^2} = |c_{0\alpha'}|^2 \N_0^2 \sum_{n,m}\sum_{e_{j},e_{k}
 \in\Gamma_{0}^{{\cal E}}}c_{0k}^{*}c_{0j}\overline{e^{-i(E_{n}-E_{m})t/\hbar}} \nonumber \\
 \times(C_{\alpha i}^{m})^{*}C_{\alpha i}^{n}C_{\alpha' k}^{m}(C_{\alpha' j}^{n})^{*}.\label{fab^2}
 \end{gather}
 As discussed in Sec.~\ref{sect-propt-chaos}, we consider the generic case that
 the spectrum of $\{E_n\}$ has no degenerate level spacing.
 This implies that $\overline{e^{-i(E_{n}-E_{m})t/\hbar}} = \delta_{nm}$.
 Then, one gets that
 \begin{gather} \notag
 \ov{|f^{(\alpha')}_{\alpha i}(t)|^2} = |c_{0\alpha'}|^2 \N_0^2 \sum_n |C_{\alpha i}^n|^2 \hspace{2cm}
 \\ \hspace{2cm} \times \Big[ \sum_{e_j,e_k \in \Gamma^{\cal E}_0} c_{0k}^* c_{0j}
 C_{\alpha' k}^n	(C_{\alpha' j}^n)^* \Big]. \label{Q-abi}
 \end{gather}

 Let us divide the summation inside the brackets on the rhs of Eq.(\ref{Q-abi})
 into two parts, one with $j \ne k$ and the other with $j=k$.
 Due to the randomness of the coefficients $c_{0j}$,
 the absolute values of the two parts have a similar order of magnitude.
 Then,  with the summation over the label $n$ performed,
 due to the irregular phases of the terms inside the brackets with $j\ne k$,
 the main contribution to the rhs of Eq.(\ref{Q-abi}) should come from the terms with $j=k$.
 Moreover, since the initial typical state $|\E_0\ra$ contains many components,
 in an approximate treatment, one may replace $|c_{0j}|^2$ in the summation by its average value
 (equal to $1$).
 Thus, finally, making using of Eq.(\ref{Q-abi}) and noting that $\N_0^2 \simeq 1/N_{\Gamma_{0}^{{\cal E}}}$,
 one gets the following simpler expression,
\footnote{ It is not difficult to verify that the rhs of Eq.(\ref{ovf-Qaai}) can also be obtained
 by an average over the initial typical state of the environment.}
\begin{gather}\label{ovf-Qaai}
 \ov{|f^{(\alpha')}_{\alpha i}(t)|^2} \simeq {|c_{0\alpha'}|^2} Q^{(\alpha')}_{\alpha i},
\end{gather}
 where, for brevity, we have introduced the following quantity,
\begin{gather}\label{Qab-i}
 Q_{\alpha i}^{(\alpha')} := \frac{1}{N_{\Gamma_{0}^{{\cal E}}}}\sum_{n}\sum_{e_{j}
 \in\Gamma_{0}^{{\cal E}}}|C_{\alpha i}^{n}|^{2}|C_{\alpha' j}^{n}|^{2}.
\end{gather}
 Then, $\overline{F}_{\alpha\alpha}^{(\alpha^{\prime}\alpha^{\prime})}$ is written as
\be\label{Faa-aa'-Q}
 \overline{F}_{\alpha\alpha}^{(\alpha^{\prime}\alpha^{\prime})}\simeq
  {|c_{0\alpha'}|^2} \sum_{i} Q^{(\alpha')}_{\alpha i} h(e_{i}).
\ee

\begin{figure}
\includegraphics[width=\linewidth]{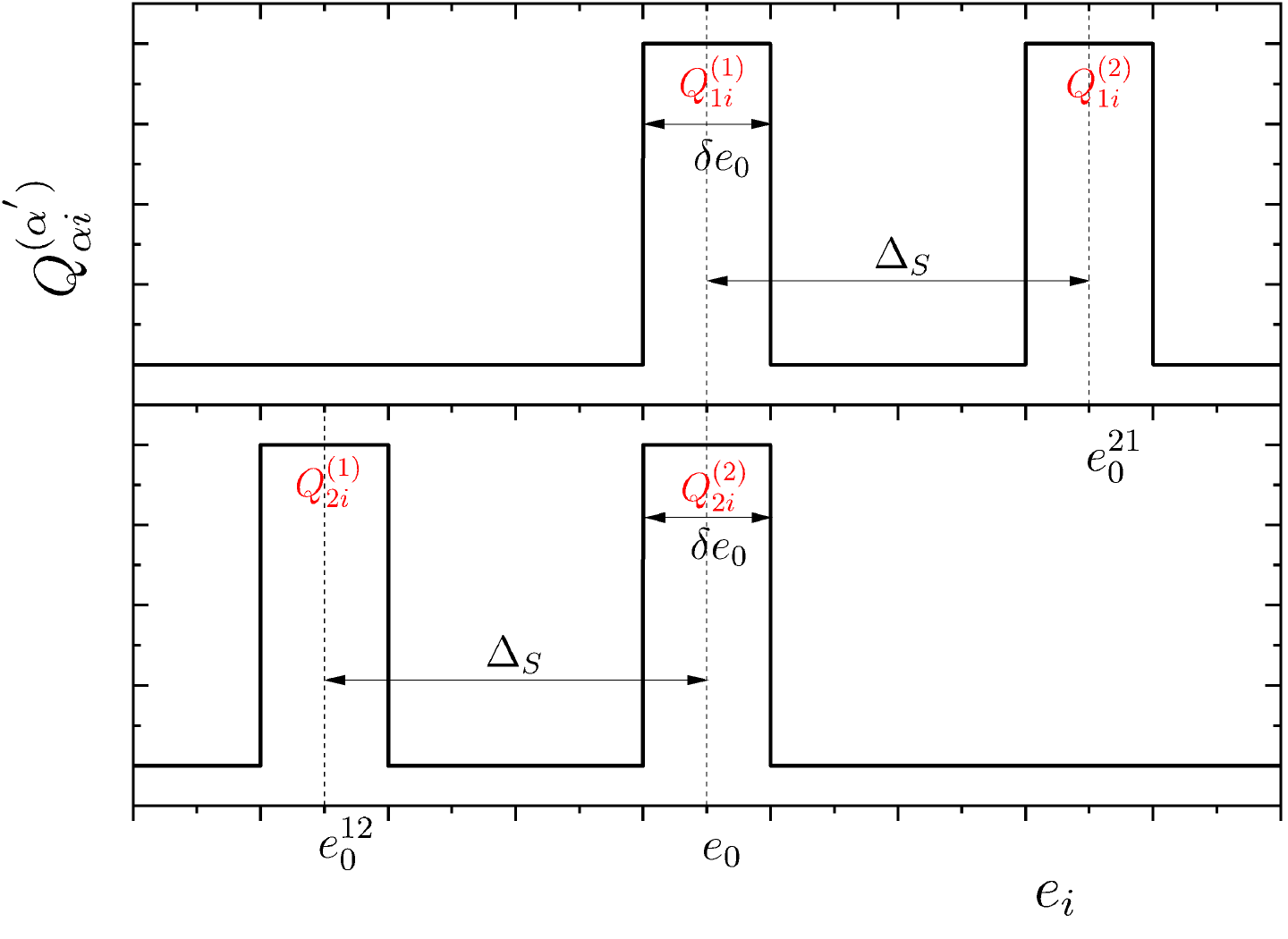}
 \caption{\label{fig:Qai}
  Schematic plots for shapes of the main-body regions of $Q^{(\alpha')}_{\alpha i}$ in Eq.(\ref{Qab-i})
  as a function of $e_i$ in the relatively strong interaction regime.
  The central region is close to the initial energy shell of $\Gamma^{\cal E}_0 = [e_0-\delta e_0/2,e_0+\delta e_0/2 ]$.
  The centers of the left and right regions are given by $e_{0}^{\alpha\alpha'} \equiv e_{0} +e_{\alpha'}^{S}-e_{\alpha}^{S}$.
 	}
\end{figure}

 Thirdly, we derive a concise expression for $\overline{F}_{\alpha\alpha}^{(\alpha'\alpha')}$.
 Making use of the assumed narrowness of the EFs of $|n\ra$ (P2), from Eq.(\ref{Qab-i}),
 one sees that only those $Q_{\alpha i}^{(\alpha')}$  for which $e^S_\alpha + e_i \simeq e^S_{\alpha'} + e_j$
 have significant values.
 This implies that the main body of $Q_{\alpha i}^{(\alpha')}$ as a function of $e_i$
 should lie within a region, which is centered at $e_{0}^{\alpha\alpha'}$ and has a width close to $\delta e_0$, where
\begin{gather}\label{}
 e_{0}^{\alpha\alpha'} \equiv e_{0} +e_{\alpha'}^{S}-e_{\alpha}^{S},
\end{gather}
 as schematically illustrated in Fig.\ref{fig:Qai}.
 Then, making use of the assumed smallness of $\delta e_0$ in P1, one finds that
\begin{gather}\label{Faaaa-Qaai}
\overline{F}_{\alpha\alpha}^{(\alpha'\alpha')}\simeq  |c_{0\alpha'}|^{2}
 h(e_{0}^{\alpha\alpha'})Q_{\alpha}^{(\alpha')},
\end{gather}
 where
 \begin{gather}\label{Q-ba}
 Q^{(\alpha')}_{\alpha } := \sum_{i} Q^{(\alpha')}_{\alpha i}.
 \end{gather}

 Fourthly, we derive a concise expression for $\overline{F}_{\alpha\alpha}$ by exploiting Eq.(\ref{Faaaa-Qaai}).
 To compute $Q^{(\alpha')}_{\alpha }$, we submit Eq.(\ref{Qab-i}) into Eq.(\ref{Q-ba})
 and write it in the following form,
\begin{gather}\label{Qab-sum-B}
Q^{(\alpha')}_{\alpha } = \frac{1}{N_{\Gamma^{\cal E}_0}} \sum_{e_j \in \Gamma^{\cal E}_0}
 \left( \sum_n |C_{\alpha' j}^n|^2  B_{\alpha }(E_{n}) \right).
\end{gather}
 Making use of the narrowness of EFs  and the slow variation of $B_\alpha (E)$ (P3),
 one gets that
\begin{gather}\label{Qab-sum-B-2}
Q^{(\alpha')}_{\alpha } \simeq \frac{1}{N_{\Gamma^{\cal E}_0}} \sum_{e_j \in \Gamma^{\cal E}_0}
 B_{\alpha }(e^S_{\alpha'} + e_j) \left( \sum_n |C_{\alpha' j}^n|^2   \right).
\end{gather}
 Noting that $\sum_n |C_{\alpha' j}^n|^2 =1$, the above expression is simplified, 
\begin{gather}\label{Qab-sum-B-2}
Q^{(\alpha')}_{\alpha } \simeq B_\alpha(e^S_{\alpha'} +e_0).
\end{gather}

 Further, from Eqs.(\ref{Faa-few}), (\ref{Faaaa-Qaai}), and (\ref{Qab-sum-B-2}), one gets that
\begin{gather}
\overline{F}_{\alpha\alpha}\simeq|c_{0\alpha}|^{2}B_{\alpha}(e_{\alpha}^{S}+e_{0})h(e_{0})
 \hspace{3.5cm} \nonumber \\
 + |c_{0\beta}|^{2}B_{\alpha}(e_{\beta}^{S}+e_{0})h(e_{0}+e_{\beta}^{S}-e_{\alpha}^{S}) \quad \ (\beta \ne \alpha).
 \label{ov-Faa-two-3}
\end{gather}
 Expanding the functions on the rhs of Eq.(\ref{ov-Faa-two-3}) linearly in the neighborhood of $e_\alpha$
and noting the fact that $|c_{0\alpha}|^2+|c_{0\beta}|^2=1$, one finds that
\begin{gather}\label{Faa-final}
\overline{F}_{\alpha\alpha}\simeq B_{\alpha}(e_{\alpha}^{S}+e_{\alpha})h(e_\alpha).
\end{gather}
 Since $\ov\rho_{\alpha\alpha}^{S}$
 can be obtained from $\overline{F}_{\alpha\alpha}$ by letting $H^{I\E}=1$
 [see Eqs.(\ref{rho-abt}) and (\ref{HIE-ab})], one finds that
\be\label{eq-rhoaa-B}
\ov\rho_{\alpha\alpha}^{S}\simeq B_{\alpha}(e_{\alpha}^{S}+e_{\alpha}).
\ee
 Finally,  one gets that
\be\label{ovrho=Qa}
\overline{F}_{\alpha\alpha}\simeq\overline{\rho}_{\alpha\alpha}^{S} h(e_\alpha).
\ee

\subsection{Proof of the second part of the main result}\label{sect-derive-main}

 In this section, we derive an expression for ${\overline F}_{12}$, then, making use of it
 and Eq.(\ref{ovrho=Qa}), we prove the second part of the main result.

 Making use of Eq.(\ref{Fab-sub}), we write  ${\overline F}_{12} =
 \sum_{\alpha', \beta'} {\overline F}_{12}^{(\alpha'\beta')}$ in the following form,
\be\label{Fab-sub-aneb}
 \overline{F}_{12} \simeq  \sum_{\alpha', \beta'}  \sum_{i}
 \overline{(f_{1 i}^{(\alpha^{\prime})})^{*}f_{2 i}^{(\beta^{\prime})}} h(e_{i}).
\ee
 Shapes of the main-body regions of $\ov{|f^{(\alpha')}_{\alpha i}(t)|^2}$
 are similar to those of $Q^{(\alpha')}_{\alpha i}$ shown in Fig.\ref{fig:Qai}.
 In an approximate treatment to the rhs of Eq.(\ref{Fab-sub-aneb}), one may consider only those terms
 for which at least one of $f_{1 i}^{(\alpha^{\prime})}$ and $f_{2 i}^{(\beta^{\prime})}$
 lies in its main-body region.
 The levels $e_i$ that contribute to these terms
 lie in three regions, which are centered at $e^{12}_0 =e_0-\Delta_S, e_0$, and $e^{21}_0=e_0+\Delta_S$,
 respectively, all with the width $\delta e_0$.

 Within the three energy regions discussed above, according to P1, $h(e_i)$ can be approximately taken as
 $h(e_0-\Delta_S), h(e_0), h(e_0 + \Delta_S)$, respectively.
 Since $h(e)$ changes slowly with $e$ and $\Delta_S$ is not large, a linear approximation is valid
 for $h(e_0 \pm \Delta_S)$, which gives
\begin{gather}\label{}
 h(e_0 \pm \Delta_S) \simeq h(e_0) \pm \Delta_S h'(e_0),
\end{gather}
 where $h'(e)$ represents the derivative of $h(e)$.
 Then, $\overline{F}_{12} $ in Eq.(\ref{Fab-sub-aneb}) is written in the following form,
\begin{gather}\label{}
 \overline{F}_{12} \simeq  h(e_{0})\overline{\rho}_{21}^{S} + G_{12} \Delta_S h'(e_0),
\end{gather}
 where
\begin{gather}\notag
 G_{12} =  \sum_{\alpha', \beta'}  {\sum_{e_i \in \Gamma^{\cal E}_0 + \Delta_s}}
 \overline{(f_{1 i}^{(\alpha^{\prime})})^{*}f_{2 i}^{(\beta^{\prime})}} \hspace{1cm}
 \\ \hspace{1cm}  -  \sum_{\alpha', \beta'}  {\sum_{e_i \in \Gamma^{\cal E}_0 - \Delta_s}}
 \overline{(f_{1 i}^{(\alpha^{\prime})})^{*}f_{2 i}^{(\beta^{\prime})}} . \label{G12}
\end{gather}
 Thus,  for $\alpha \ne \beta$ one gets that
\be\label{Fab-final}
\overline{F}_{\alpha\beta}\simeq h(e_{0})\overline{\rho}_{\beta\alpha}^{S} + G_{\alpha \beta} \Delta_S h'(e_0).
\ee

 Substituting Eqs.(\ref{ovrho=Qa}) and (\ref{Fab-final}) into  Eq.(\ref{tls-rho}),
 one gets that
\begin{gather}\label{main-pre}
 \overline{\rho}_{\alpha\beta}^S \simeq \frac{ \eta_r
 (h_\beta\ov{\rho}_{\beta\beta}^S-h_\alpha\ov{\rho}_{\alpha\alpha}^S) }{1-\eta_d h_0}
 + \frac{ \eta_d G_{\beta\alpha } \Delta_S h'(e_0)}{1-\eta_d h_0} .
\end{gather}
 For a large environment, usually $h'(e)$ scales as $1/N$ \cite{huse2014,rigol2009breakdown}.
 Hence,  for a sufficiently large environment, one has
\begin{gather}\label{HIS-h'<<1}
  |(H_{\alpha\alpha}^{IS}-H_{\beta\beta}^{IS}) h'(e_0)| \ll 1.
\end{gather}
 This gives that
\begin{gather}\label{etarhohp}
 |  \eta_d \ov{\rho}_{\alpha\beta}^S \Delta_S h'(e_0)| \ll |\ov{\rho}_{\alpha\beta}^S|.
\end{gather}
 Moreover, since $G_{\beta\alpha}$ in Eq.(\ref{G12}) contains less terms than $\ov\rho^S_{\alpha\beta}$,
 one usually has $|G_{\beta\alpha}| \lesssim |\ov\rho^S_{\alpha\beta}|$.
 In the relatively strong regime of interaction, $|\eta_d h_0|$ is considerably smaller than $ 1$.
 Due to these properties and Eq.(\ref{etarhohp}),
the absolute value of the second item on the rhs of Eq.(\ref{main-pre})
 is much smaller than $|\ov\rho^S_{\alpha\beta}|$; hence, this term can be neglected.
 This finishes the proof of Eq.(\ref{main-result}).

\section{ Pure-dephasing and very-weak dissipative interactions} \label{sect-derive-pd-vw}

 In this section, we discuss  $\overline{\rho}_{\alpha\beta}^S$
 under two special types of the system-environment interaction:
 one being  very weak dissipative interactions
 and the other  being energy-preserving interactions that may induce pure dephasing.

 \begin{figure}
	\includegraphics[width=\linewidth]{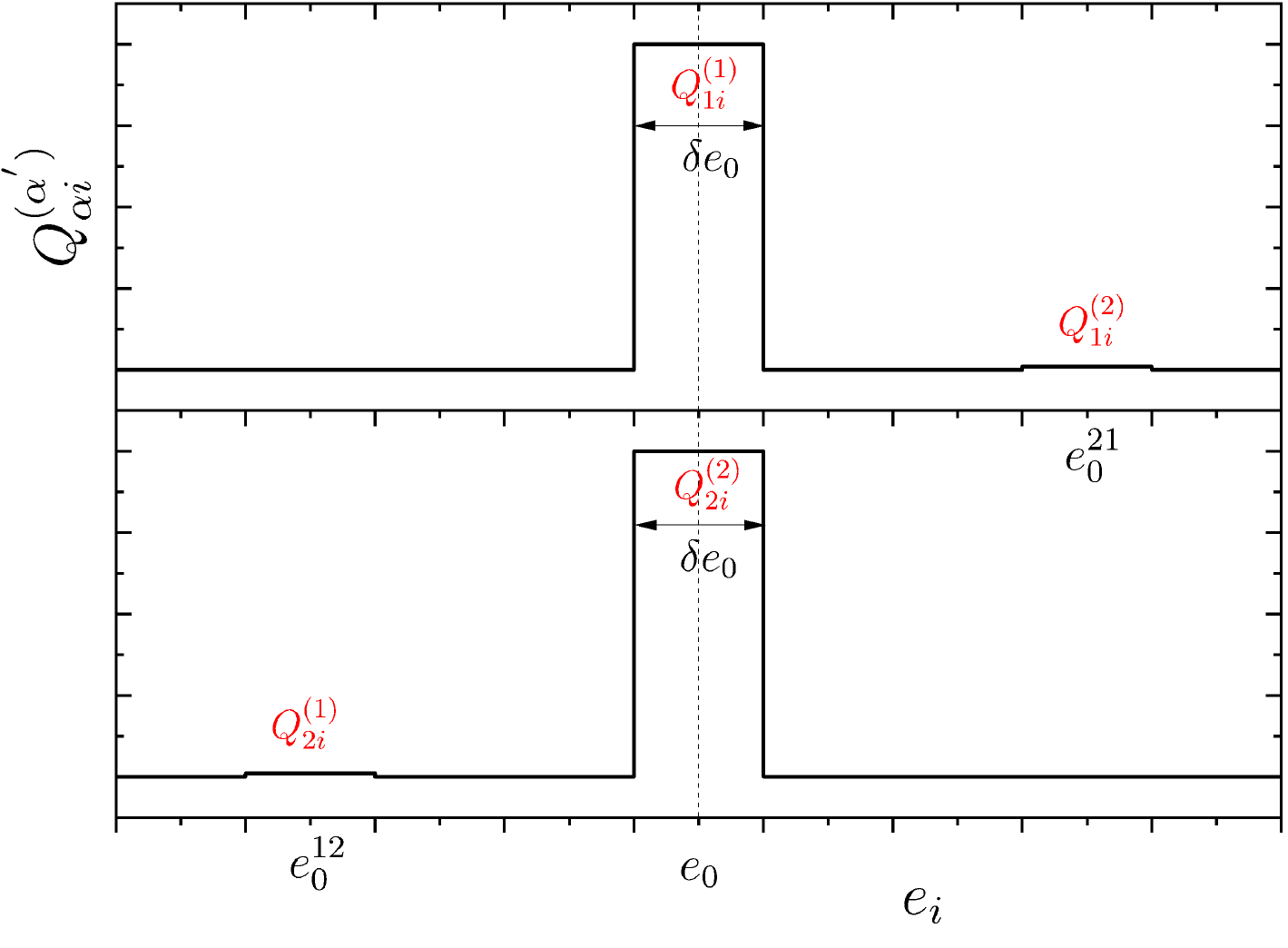}
	\caption{\label{fig:Qai0}
  Similar to Fig.\ref{fig:Qai}, but for an very weak  interaction.
	}
\end{figure}

\vspace{0.2cm}
\noindent \emph{1. Very weak dissipative interactions}.
 \\ Under such an interaction, each eigenstate $|n\ra$ has only one significant component
 when being expanded in the basis of $|E_r\ra$.
 Thus, there is one label denoted by $r(n)$, such that $|C^n_{r(n)}| \simeq 1$.
 As a consequence, the value of $B_\alpha (E_n)$ fluctuate nearly between $0$ and $1$ with the change of $E_n$.
 This invalidates those discussions given in Sec.~\ref{sect-Faa} that are after Eq.(\ref{Qab-sum-B}),
 but, discussions before Eq.(\ref{Qab-sum-B}) are still valid.

 Making use of the property of $|C^n_{r(n)}| \simeq 1$, from Eq.(\ref{Qab-i}), one finds that
\begin{gather}\label{Qab-i-ew}
 Q_{\alpha i}^{(\alpha')} \simeq \frac{1}{N_{\Gamma_{0}^{{\cal E}}}}
 \left\{
   \begin{array}{ll}
     \delta_{\alpha \alpha'} , & \hbox{if $e_{i} \in\Gamma_{0}^{{\cal E}}$;} \\
     0, & \hbox{otherwise.}
   \end{array} \right.
\end{gather}
 This implies that $Q^{(1)}_{1}\simeq Q^{(2)}_2\simeq 1$, while, $Q^{(2)}_{1}\simeq Q^{(1)}_2\simeq 0$,
 as illustrated in Fig.\ref{fig:Qai0}.
 In this case, Eq.(\ref{Faa-few}) becomes
\begin{gather}\label{Fab0}
  \overline{F}_{\alpha\alpha}\simeq 	\overline{F}_{\alpha\alpha}^{\alpha\alpha}.
\end{gather}
 Then, using Eq.(\ref{Faa-aa'-Q}), one finds that $\overline{F}_{\alpha\alpha}^{\alpha\alpha}
\simeq |c_{0\alpha}|^2 h_0$.
 Moreover, one notes that $|\eta_d h_0| \ll 1$  in this interaction regime and,
 following some arguments similar to those used in Sec.~\ref{sect-derive-main},
 one finds that the contribution from $\overline{F}_{\alpha\beta}$ with $\beta \ne \alpha$ can be neglected.
 Finally, making use of Eq.(\ref{tls-rho}), one finds that
\begin{gather} \label{tls-relax-weak-gen}
\overline{\rho}_{\alpha\beta}^S \simeq \eta_r h_0 ( |c_{0\beta}|^2 - |c_{0\alpha}|^2)
\quad (\beta \ne \alpha).
\end{gather}

\vspace{0.2cm}
\noindent \emph{2. Energy-preserving interactions}.
\\ An energy-preserving interaction
 satisfies $[H^S,H^{IS}]=0$, implying that $H^{IS}_{\alpha\beta}=0$ for $\alpha \ne \beta$.
 Such an interaction induces the well-known effect of pure dephasing.
 Here, it is more convenient to study $\overline{\rho}_{\alpha\beta}^S$ with $\alpha \ne \beta$
 in a direct way \cite{gorin2004connection}, not using Eq.(\ref{tls-rho}).

 In fact, under such an  interaction,  according to Eq.\eqref{alpha-t}, each branch $|\E_\alpha(t)\rangle$
 follows a Schr\"{o}dinger evolution under the effective Hamiltonian $H^{\rm eff}_{\E\alpha}$,
\begin{gather}\label{Ea-time-evol}
 |\E_\alpha(t)\rangle = e^{-it H^{\rm eff}_{\E\alpha}/\hbar}|\E_0\ra.
\end{gather}
 Substituting this expression into Eq.(\ref{rho-abt}), one gets that
\begin{gather}\label{rho-LE}
\rho_{\alpha\beta}^S(t) = \la \E_0| \exp\left( it H^{\rm eff}_{\E\beta}/\hbar \right)
\exp\left( -it H^{\rm eff}_{\E\alpha}/\hbar \right)|\E_0\ra.
\end{gather}
 The rhs of Eq.(\ref{rho-LE}) has a form like that of the so-called quantum Loschmidt echo (LE) amplitude.
 It is known that  the LE decays quickly in quantum chaotic systems, usually in an exponential way
 \cite{gorin2006dynamics,prosen2002stability,cerruti2002sensitivity,
 fid-chao-JP,fid-chao-Bee,fid-chao-Cuc,wang-LE1,wang-LE2},
 implying that the two branches $|\E_\alpha(t)\rangle$ separate fast.
 When the environment can be modeled by random matrices and
 the system-environment interaction can significantly disturb the environment,
 for long times, the relation between $|\E_\beta(t)\ra $ and $|\E_\alpha(t)\rangle$ is effectively
 like that between two randomly chosen state vectors in the Hilbert space of the environment,
 which implies that $\rho_{\alpha\beta}^S(t) \sim d^{-1/2}_\E$
  \cite{prosen2002stability,gorin2006dynamics,foot-puredephas}.

\begin{figure}
	\includegraphics[width=\linewidth]{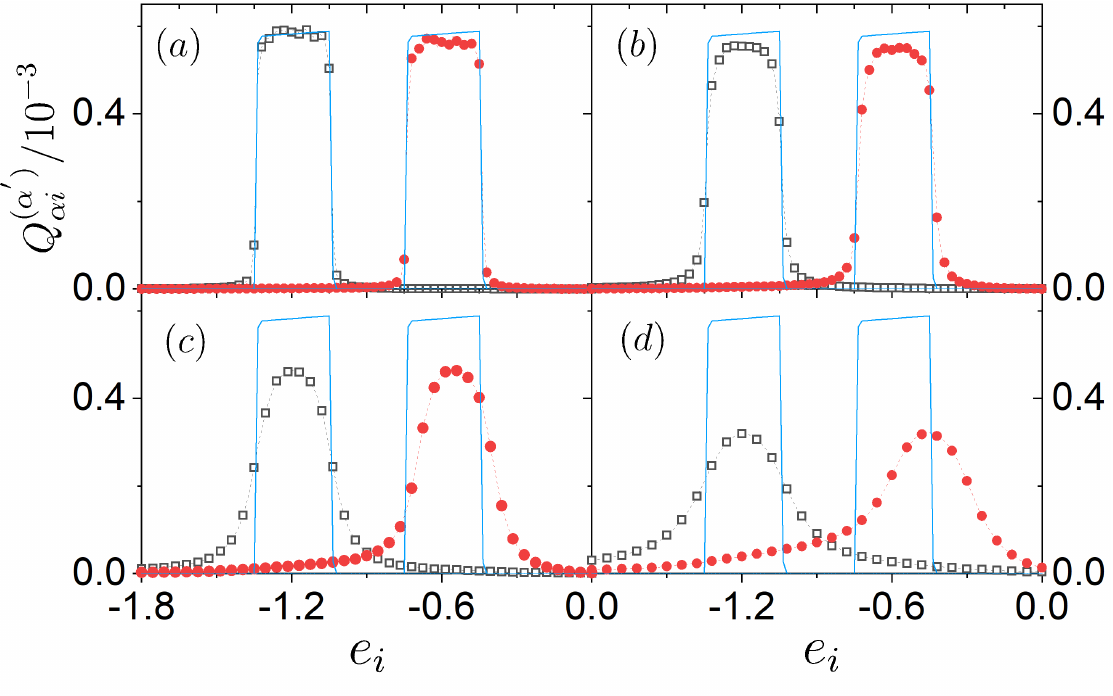}
	\caption{\label{fig:cn} Coarse-grained shapes of $Q^{(1)}_{1i}$ (open squares) and $Q^{(2)}_{1i}$ (solid circles)
	 as functions of $e_i$.
		Parameters: $N=14$, $\Delta_S = 0.6$, and $\delta e_0 =0.3$.
 (a) $\lambda = 0.15$, (b) $\lambda =0.3$, (c) $\lambda= 0.6$, and (d) $\lambda =1.0$.
	}
\end{figure}

 To compute the long-time average of $\langle \E_\beta(t)|\E_\alpha(t)\rangle$, one may
 expand the initial state $|\E_0\ra$ as
 \be\label{eq-initialsp}
|{\cal E}_{0}\rangle=\sum_k D_{\alpha k}|\psi_{k}^{\alpha}\rangle=\sum_k D_{\beta l}|\psi_{l}^{\beta}\rangle,
 \ee
 where, $|\psi_{k}^{\alpha}\rangle$ and $|\psi_{k}^{\beta}\rangle$ indicate the eigenstates of
 $H^{\rm eff}_{\E\alpha}$ and $H^{\rm eff}_{\E\beta}$, respectively,
 \be\label{eq-Heff}
\ensuremath{H_{{\cal E}\alpha}^{{\rm eff}}|\psi_{k}^{\alpha}\rangle={\varepsilon}_{k}^{\alpha}
|\psi_{k}^{\alpha}\rangle.}
 \ee
 Using Eqs.(\ref{eq-initialsp})-(\ref{eq-Heff}),
 $\overline{\langle{\cal E}_{\beta}(t)|{\cal E}_{\alpha}(t)\rangle}$ is written as
 \be
 \overline{\langle{\cal E}_{\beta}(t)|{\cal E}_{\alpha}(t)\rangle}
 =\sum_{k,l}D_{\alpha k}D_{\beta l}^{*}\overline{\exp(-i({\varepsilon}_{k}^{\alpha}-{\varepsilon}_{l}^{\beta})t)}
 \langle\psi_{l}^{\beta}|\psi_{k}^{\alpha}\rangle.
 \ee
 If $\varepsilon_{k}^{\alpha}  \ne \varepsilon_{l}^{\beta}$ for all $k$ and $l$ with $\beta\ne \alpha$, one gets that
 \be
 \overline{\exp(-i({\varepsilon}_{k}^{\alpha}-{\varepsilon}_{l}^{\beta})t)}=0, \quad \forall k,l.
 \ee
 This gives that $\ov{\langle \E_\beta(t)|\E_\alpha(t)\rangle} =0$.
 Hence, $\overline{\rho}_{\alpha\beta}^S =0$ for $\alpha \ne \beta$.

\section{Numerical Simulations}\label{sect-numerical}

 In this section, we discuss numerical simulations that have been performed
 for the purpose of testing analytical predictions given above.
 We first discuss the employed model in Sec.~\ref{sect-model}.
 Then,  in Sec.~\ref{sect-numeric-ETH},
 we give numerical evidences that the ETH ansatz is applicable to the environment system in the employed model.
 Finally,  in Sec.~\ref{sect-numeric-mr}, we discuss numerical simulations for testing the predicted relations
 among elements of $\ov\rho^S$.

\subsection{The model employed}\label{sect-model}

 We employ a model that consists of one qubit as the central system $S$
 and one defect Ising chain as the environment $\E$.
 The Ising chain is composed of $N$ $\frac{1}{2}$-spins
 lying in a nonhomogenous transverse field, whose Hamiltonian is written as
\begin{equation}
H^\E = h_x\sum_{l=1}^NS_l^x + d_1 S_1^z + d_5 S_5^z  +  J_z\sum_{l=1}^NS_l^z S^z_{l+1},
\end{equation}
 where $S^x_l$ and $S^z_l$ indicate Pauli matrices divided by $2$ for the $l$-th site.
 The parameters $h_x$, $J_z$, $d_1$ and $d_5$ are adjusted,
 such that the defect Ising chain is a quantum chaotic system.
 That is, for levels not close to edges of the energy spectrum,
 the nearest-level-spacing distribution $P(s)$
 is close to the Wigner-Dyson distribution $P_W(s) = \frac{\pi}{2}s\exp(-\frac{\pi}{4}s^2)$,
 the latter of which is almost identical to the prediction of the RMT.
 The exact values of the parameters used are $h_x = 0.9, J_z = 1.0$,  $d_1 = 1.11$, and $ d_5 = 0.6$.
 In our numerical simulations, the periodic boundary condition was used.

\begin{figure}
	\includegraphics[width=1.0\linewidth]{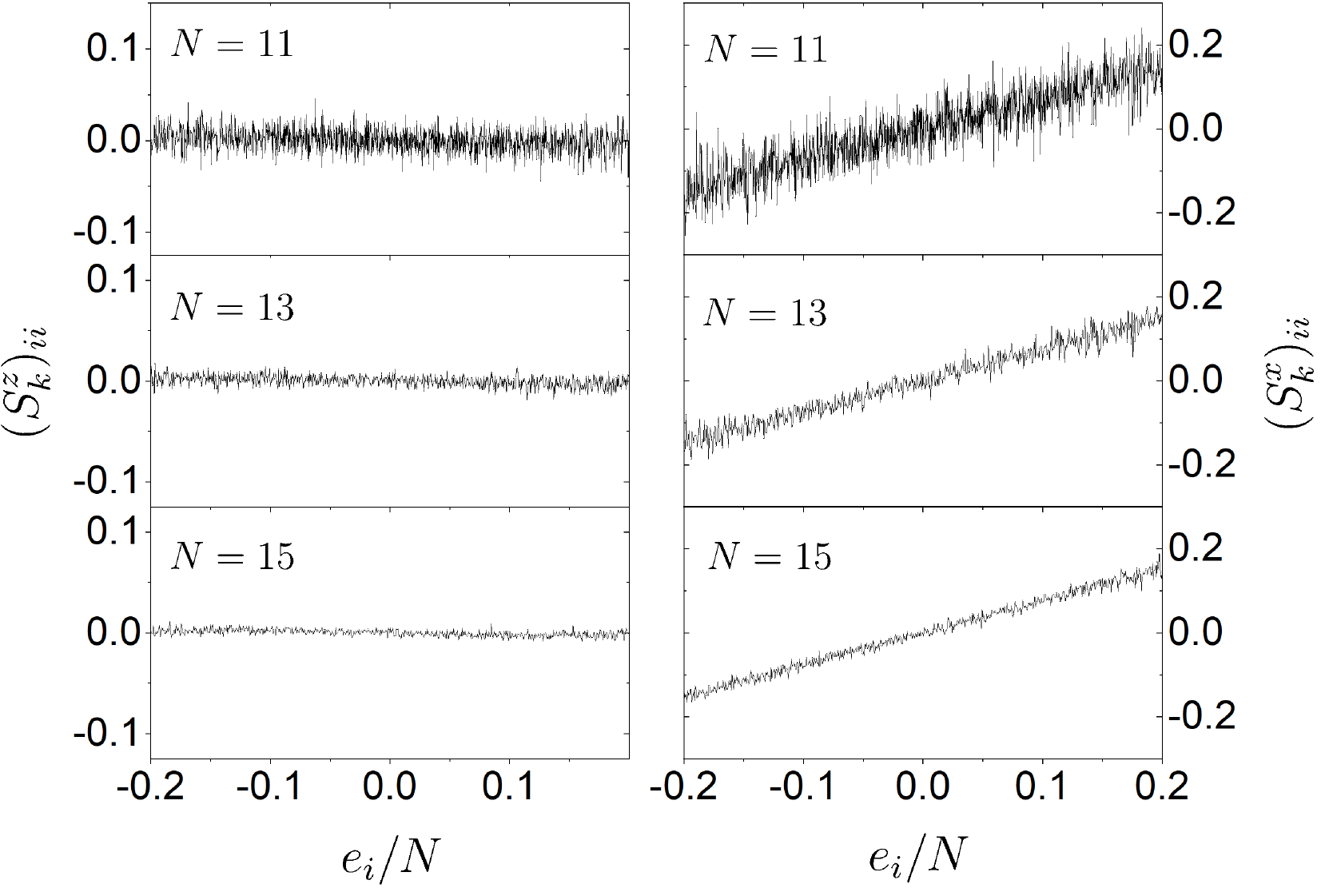}
	\caption{\label{fig:diag} Dependence of diagonal elements of two local observables
		$S_k^z$ and $S_k^x$ of the defect Ising chain on the environmental eigenenergy $e_i$ ($k=7$).
		In agreement with the ETH ansatz in Eq.\eqref{ETH}, these elements fluctuate around
		certain slowly-varying functions of $e$, respectively, and the fluctuations decrease with the increase of the
particle number $N$.
	}
\end{figure}

 The central system is coupled to the $k$-th spin of the Ising chain, with a self-Hamiltonian
\begin{gather}\label{}
 H^S = \Delta_S S^z.
\end{gather}
 In numerical simulations, we set $\Delta_S =0.6$.
 Four types of the interaction Hamiltonian $H^I$ have been studied, denoted by $H^I_{(\kappa)}$
 with $\kappa =1,2,3,4 $.
 Specifically, the first two types are written as
\begin{gather}\label{}
  H^I_{(1)} =\lambda S^x\otimes S^z_k,
 \\ H^I_{(2)} =\lambda S^x\otimes S^x_k,
\end{gather}
 with a difference lying in the environmental part $H^{I\E}$.
 Both of them describe dissipative interactions, with $H^{IS}_{\alpha\alpha}=0$.
 Numerically, we found that the function $h(e)$ for the first type ($H^I_{(1)}$)
 has quite small values, such that the requirement in Eq.(\ref{hc>>HIE}) is not satisfied.
 In contrast, the values of $h(e)$ for the second type ($H^I_{(2)}$) are not small, satisfying Eq.(\ref{hc>>HIE}).
 Thus, according to discussions given in Sec.~\ref{sect-main},
 Eq.(\ref{main-result}) should work better in the case of $H^I_{(2)}$ than for  $H^I_{(1)}$.

\begin{figure}
	\includegraphics[width=0.95\linewidth]{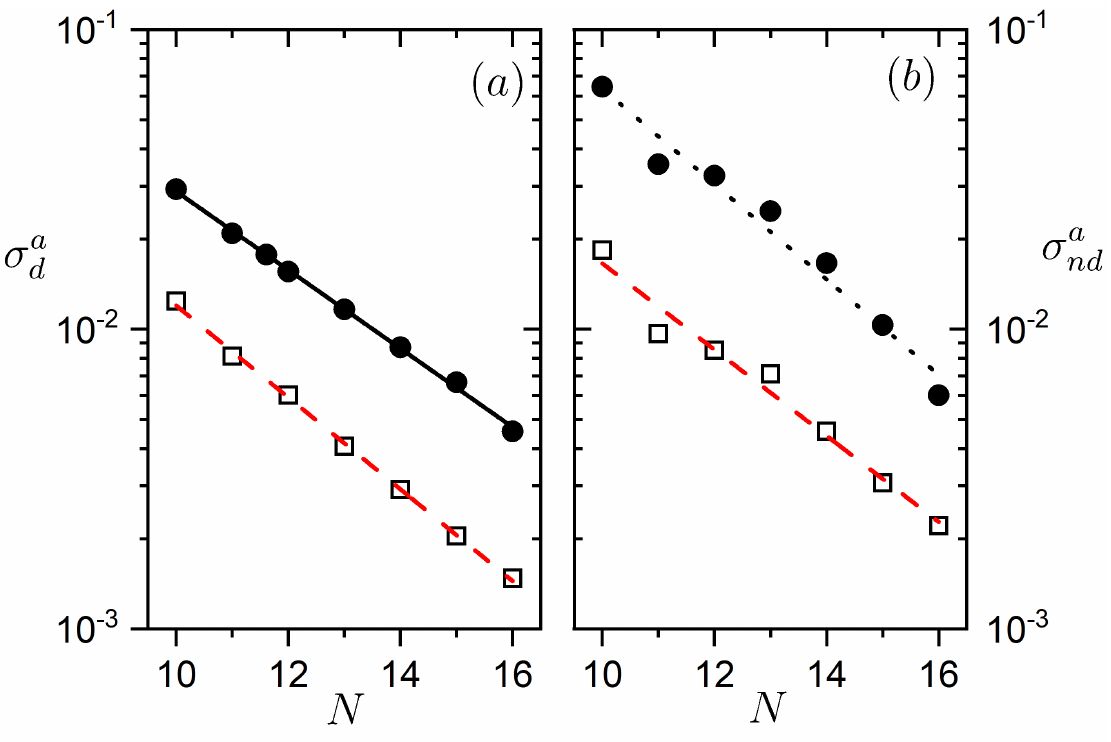}
	\caption{\label{fig:fluc} (a) Exponential decay of the deviation $\sigma_d^a$ in Eq.\eqref{sigma-da}
		with the increase of $N$,
		for fluctuations of the diagonal elements of $S_k^x$ (empty squares) and of $S_k^z$ (solid circles).
		(b) Exponential decay of $\sigma^a_{nd}$ in Eq.\eqref{sigma-nda} for fluctuations of offdiagonal elements.
		The results are in agreement with the prediction of ETH in Eq.\eqref{ETH}.
}
\end{figure}

 The third type is for a dissipative interaction with a more generic form in the central system's part, written as
\begin{gather}\label{}
 H^I_{(3)} = \lambda(S^z +S^x)\otimes S_k^x.
\end{gather}
 The fourth type gives a pure-dephasing interaction, 
\begin{gather}\label{pd}
 H^I_{(4)} = \lambda S^z\otimes S^z_k,
\end{gather}
 which is commutable with $H^S$.

 In numerical simulations, the values of $\ov{\rho}^S_{\alpha\beta}$ were computed by taking the
 long-time average over the following expression of ${\rho}^S_{\alpha\beta}$,
\begin{equation}
\rho^S_{\alpha\beta}(t) = \sum_{nm,i}e^{-i(E_{n}-E_m)t}\langle\alpha i|n\rangle\langle n
|\Psi(0)\rangle\langle\Psi(0)|m\rangle\langle m|\beta i\rangle.
\end{equation}
 Since the eigenenergies $E_{n}$ have no degeneracy, practically, we computed
\begin{equation}\label{rho-ab-comp}
\ov{\rho}^S_{\alpha\beta}
=\sum_{i,n}|\langle n|\Psi(0)\rangle|^2 C_{\beta i}^{n*} C_{\alpha i}^n .
\end{equation}

 To give a comparison with the schematic plots in Fig.\ref{fig:Qai}, some examples of the 
 exact (coarse-grained) shapes of $Q^{(\alpha')}_{\alpha i}$  are shown in Fig.\ref{fig:cn}.
 In each panel, the solid lines in the left part indicate the coarse-grained shape obtained in 
 the very weak case of Eq.(\ref{Qab-i-ew}) with $\alpha'=\alpha$; those in the right 
 part are obtained just by a shift in the energy by $\Delta_S$. 
 We found that, under interaction not strong ($\lambda \lesssim 0.15$), 
 the exact shapes are close to the schematic plots in the upper panel of Fig.\ref{fig:Qai}.
 With further increase of the interaction strength, obvious deviations appear.

\begin{figure}
\includegraphics[width=1.0\linewidth]{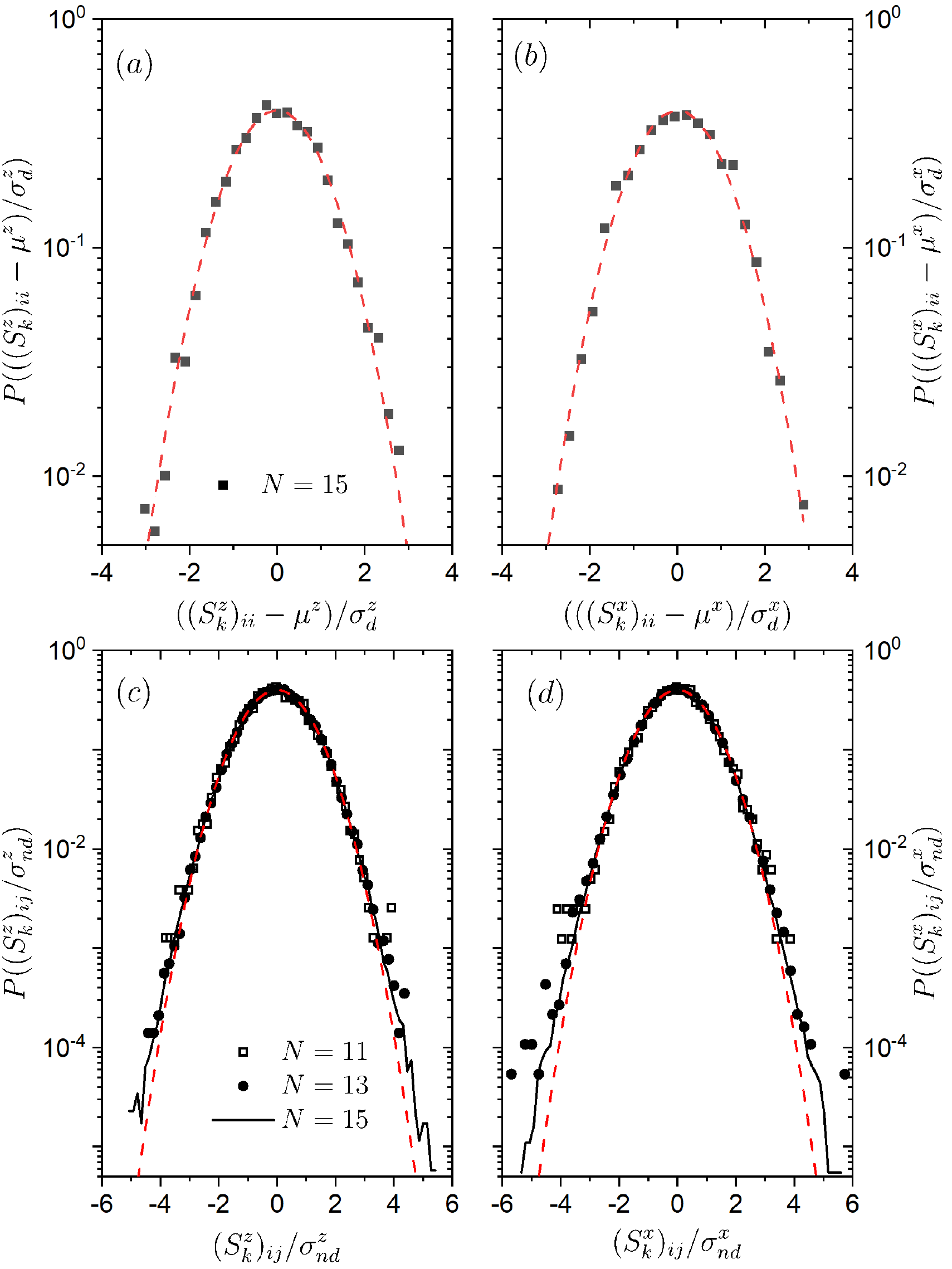}
 \caption{\label{fig:nondiag} Distributions of fluctuations of the diagonal elements of  $S_k^z$
 	[(a)] and of $S_k^x$ [(b)], rescaled by $\sigma_{d}^z$ and $\sigma_{d}^x$, respectively.
 And, distributions of the offdiagonal
 elements of $S_k^z$ [(c)] and of $S_k^x$ [(d)], rescaled by $\sigma_{nd}^z$ and $\sigma_{nd}^x$, respectively. 
 The dashed curves represent the Gaussian distribution with unit variance.
 }
\end{figure}

\subsection{Numerical tests of ETH}\label{sect-numeric-ETH}

 Due to the hypothesis nature of the ETH ansatz in Eq.\eqref{ETH},
 we have tested its validity in the model employed in this paper.
 We did this for the two local operators $S_k^x$ and $S_k^z$ at the site $k=7$ in the defect Ising chain.

 The so-called Krylov-space method was employed in our numerical computation of
 the EFs, which are around the energy of $e_{0} = -1.2$ in the middle energy region.
 For example, we calculated eigenstates for $N=11,13,15$ with $\delta e_{0} \simeq 0.3$.

 Let us first discuss predictions of Eq.\eqref{ETH} for diagonal elements of local observables.
 Expectation values of the two local observables, i.e.,
\begin{gather}\label{Skx-ij}
 (S_k^a)_{ii} = \la i |S_k^a|i\ra \quad \text{with $a=x,z$,}
\end{gather}
 are plotted  in Fig \ref{fig:diag}.
 It is seen that, in agreement with ETH, the diagonal elements fluctuate around
 certain slowly-varying function $h(e)$ and the fluctuation decreases with the increase of the particle number $N$.
 Note that the horizontal axis is labelled with $e_i/N$.
 For $a=z$, the values of $h(e)$ are close to zero, while, for $a=x$, most of $|h(e)|$ are notably larger than zero.

 To study quantitatively the fluctuations of $(S_k^a)_{ii}$,
 we have computed the standard deviations  $\sigma^a_d$,
 \begin{equation}\label{sigma-da}
\sigma_d^a = \sqrt{\frac{1}{N_{\Gamma_0^\E}}\sum_{e_i\in\Gamma_0^\E}|(S_k^a)_{ii}-\mu^a|^2},
\end{equation}
where
\begin{equation}
\mu^a = \frac{1}{N_{\Gamma_0^\E}}\sum_{e_i\in\Gamma_0^\E} (S_k^a)_{ii}.
\end{equation}
 As seen in Fig.~\ref{fig:fluc} (a), the fluctuation decays exponentially with the increase of $N$,
 as predicted by the term $e^{-S(e)}$ in the second part on the rhs of Eq.\eqref{ETH}.
 Moreover, in agreement with the prediction of ETH, the distributions of $[(S_k^a)_{ii} - \mu^a]/\sigma^a_{d}$
 are close to the Gaussian form [Fig.~\ref{fig:nondiag} (a) and (b)].

 Next, we discuss the offdiagonal elements $(S_k^a)_{ij}$.
 In agreement with the prediction of ETH, the probability distributions of $(S_k^a)_{ij}/\sigma^a_{nd}$
 have a Gaussian form [Fig.~\ref{fig:nondiag} (c) and (d)], where $\sigma^a_{nd}$ are the standard deviations
 for the offdiagonal elements,
 \begin{equation}\label{sigma-nda}
\sigma_{nd}^a = \sqrt{\frac{1}{N_{\Gamma_0^\E}(N_{\Gamma_0^\E}-1)}\sum_{i\ne j\in\Gamma_0^\E}|(S_k^a)_{ij}|^2}.
\end{equation}
 These standard deviations also decay exponentially with the increase of $N$ [Fig.~\ref{fig:fluc} (b)].

\begin{figure}
\includegraphics[width=\linewidth]{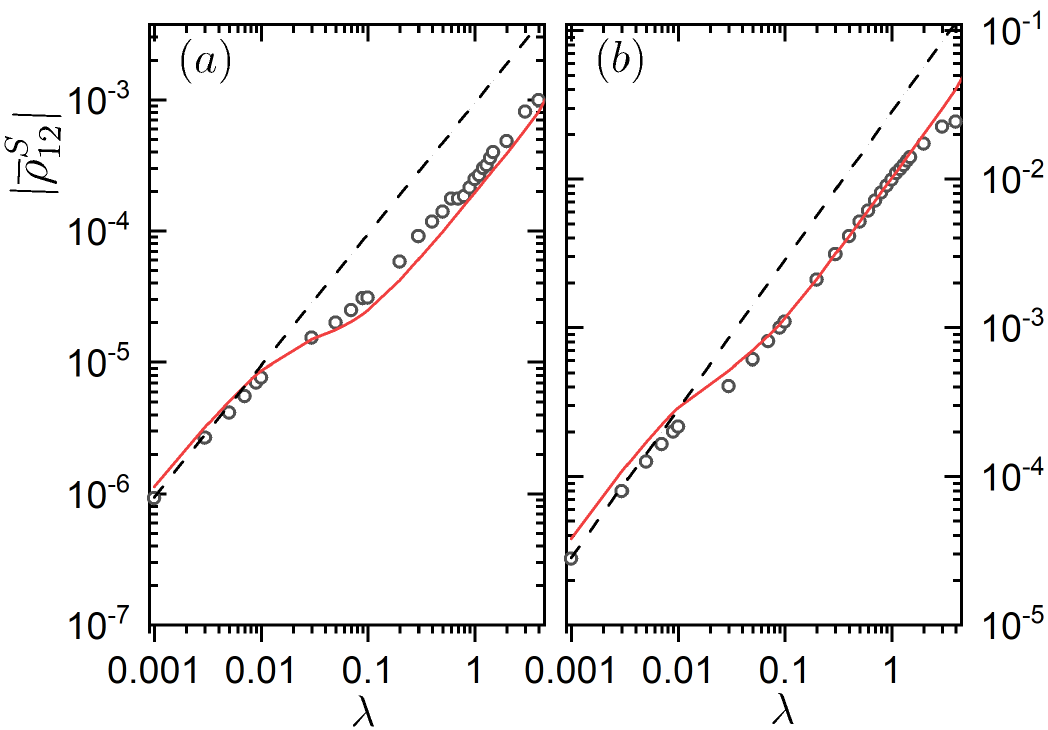}
\caption{\label{fig:lta-sx-all} The values of $|\ov\rho^S_{12}|$ (circles)
versus the coupling strength $\lambda$ in the logarithm scale,
 under interactions (a) $H^I_{(1)} =\lambda S^x\otimes S^z_k$ and (b) $H^I_{(2)} =\lambda S^x\otimes S^x_k$.
 The initial state of the system $S$ is a superposition state with $(c_{01}, c_{02})=(1/2,\sqrt{3}/2)$.
 The solid lines indicate predications of Eq.\eqref{main-result}
 for relatively-strong interactions, while,
 the dashed lines represent predications of Eq.(\ref{tls-relax-weak-gen}) for very weak interactions.
 Parameters: $N=13$, $e_0=-1.2$.
 }
\end{figure}

\subsection{Numerical tests for the main results}\label{sect-numeric-mr}

\begin{figure}
\includegraphics[width=\linewidth]{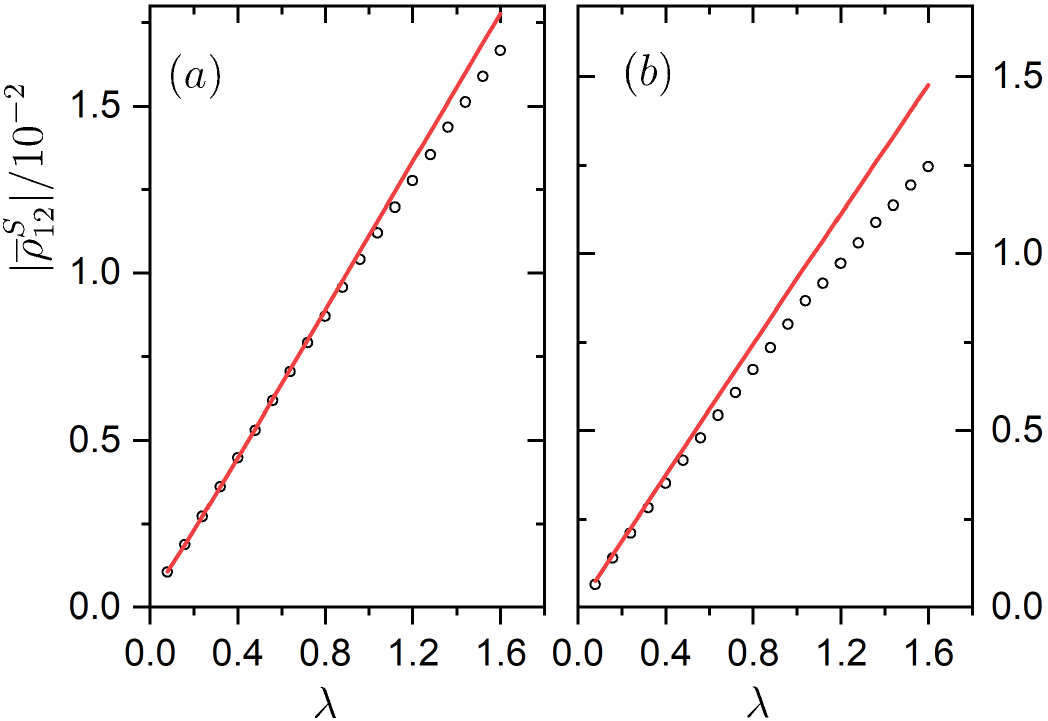}
\caption{\label{fig:lta-nd-sx} $|\ov\rho^S_{12}|$
 versus $\lambda$ in the normal scale for $H^I_{(2)} =\lambda S^x\otimes S^x_k$,
 (a) $(c_{01},c_{02})=(0,1)$ and (b) $(c_{01},c_{02})=(1/\sqrt{2},1/\sqrt{2})$.
 The solid lines are predictions of  Eq.\eqref{main-result}.
}
\end{figure}

 In this section, we discuss numerical simulations that have been performed
 in order to test analytical predictions given in previous sections.

 Since the second part of the main result of this paper is for a situation that is more complicated and generic than the
 first part, we present numerical tests for the second part only.
 This second part predicts that, under certain conditions, elements of $\ov\rho^S$ should approximately satisfy
 Eq.(\ref{main-result}) for interactions lying in the relatively strong regime.
 Some examples of our numerical simulations are given in Fig.\ref{fig:lta-sx-all},
 showing variation of $|\ov\rho^S_{12}|$ versus the interaction strength $\lambda$
 in the logarithm scale.
 In the computation of the prediction of Eq.(\ref{main-result}) for the offdiagonal element $|\ov\rho^S_{12}|$,
 exact values of the diagonal elements $\ov\rho^S_{\alpha\alpha}$ on the rhs of the equation were used,
 which were obtained directly from numerical simulations with Eq.(\ref{rho-ab-comp}).

 A wide scope of the interaction strength is plotted in Fig.\ref{fig:lta-sx-all}, from the very weak regime
 ($\lambda \lesssim 0.005$)  to the very strong regime ($\lambda \gtrsim 1)$.
 As mentioned above, Eq.(\ref{main-result}) is expected to be valid for the relatively-strong
 regime, which correspond to the middle region in Fig.\ref{fig:lta-sx-all}.
 The right panel shows that, under the interaction $H^I_{(2)}$,
 the analytical predictions are close to the exact results in the middle region.

 However, in the left panel  of Fig.\ref{fig:lta-sx-all} for $H^I_{(1)}$,
 the predictions of Eq.(\ref{main-result}) are notably below the exact values in the middle region,
 though giving the correct trend. 
 This deviation is in fact due to the fact that Eq.(\ref{hc>>HIE}), as a prerequisite of
 Eq.(\ref{main-result}), is not satisfied.
 Concretely, for $H^I_{(1)}$ with $ H^{I\E_1} = S^z_k$, 
 we found that $\la |h|\ra  \simeq 0.0023$ and $\la H^{I\E}_{\rm fluc} \ra \simeq 0.0041$.
 In this case, the contribution from the terms $ \Delta_{\alpha\beta}$ to $|\ov\rho^S_{12}|$, 
 which have been neglected in the derivation of Eq.(\ref{main-result}), 
 should be nonnegligible.
 This explains the phenomenon that the predictions of Eq.(\ref{main-result})
 are notably smaller than the exact values of $|\ov\rho^S_{12}|$.
 In contrast, for $H^I_{(2)}$, we found that Eq.(\ref{hc>>HIE}) is valid,
 with $\la |h|\ra  \simeq 0.07$ much larger than $\la H^{I\E}_{\rm fluc} \ra  \simeq 0.01$.

\begin{figure}
\includegraphics[width=\linewidth]{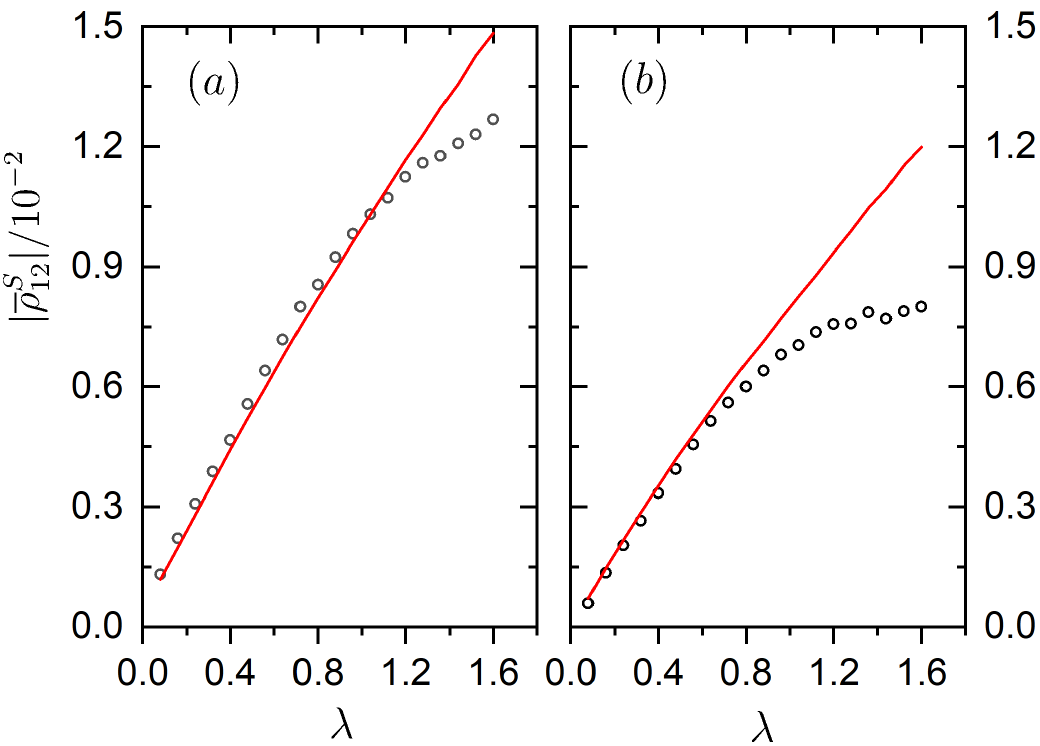}
\caption{\label{fig:lta-mix}Similar to Fig.\ref{fig:lta-nd-sx}, but for
 $H^I_{(3)} = \lambda(S^z +S^x)\otimes S_k^x$.
}
\end{figure}

 Interestingly, we found that Eq.(\ref{main-result}) is applicable even for some 
 quite strong interactions.
 Some examples are shown in Fig.\ref{fig:lta-nd-sx} and Fig.\ref{fig:lta-mix}.
 For $H^I_{(2)}$, the norm $\| H^{I}_L \|$ in Eq.(\ref{norm-HI}) is equal to $\lambda /2$ 
 and, hence, the parameter values of $\lambda$ around $1$ belong to the regime of very strong interaction.

 Moreover, Eq.(\ref{tls-relax-weak-gen}) works quite well in the very weak interaction regime
 in both panels of Fig.\ref{fig:lta-sx-all}.
 In fact, making use of the fact that each EF contains only one big component in the very weak
 interaction regime, it is not difficult to find that the contributions of $\Delta_{\alpha\beta}$ are always small,
 independent of whether Eq.(\ref{hc>>HIE}) holds.
 As for the pure-dephasing interaction Hamiltonian $H^I_{(4)}$,
 numerically, we found that $\ov{\rho}^S_{\alpha\beta}$ are indeed equal to zero as predicted analytically.

\section{Conclusions and discussions}\label{sect-conclusion}

 To summarize, we have studied the long-time averaged RDM $\overline{\rho}^S$ of a TLS (qubit),
 which is locally coupled to a generic many-body  quantum chaotic system (as the environment),
 with the total system undergoing a Schr\"{o}dinger evolution.
 Approximate relations among elements of $\ov\rho^S$ in the eigenbasis of the TLS
 have been derived in several situations, 
 particularly for generic dissipative interactions in the relatively strong regime.

 The derived relations may find applications in the study of many topics, some of which we discuss below.
 Firstly, they are useful in the study of steady-state properties.
 Compared with results obtained from the master-equation approach, which is based on
 a perturbative treatment and whose validity for long times is a subtle issue,
 results obtained from the approach employed in this paper
 should be more reliable for long times, though at the cost of losing details of the evolution.

 Some further comments for steady states:
 For systems of the type studied in this paper, it is already known that
 the RDM $\rho^S(t)$ is close to $\ov\rho^S$ for most of the times in the long-time limit,
 with fluctuations scaling as $N_{\Gamma^{\cal E}_0}^{- 1/2}$
\cite{von2010proof,linden2009quantum, reimann2008foundation,
	reimann2015eigenstate,reimann2012equilibration,short2011equilibration,short2012quantum}.
 Moreover, although analytical demonstration of the emergence of steady states is a subtle topic,
 numerical simulations show that such states may emerge in many situations.

 Secondly, the derived relations impose approximate restrictions to the freedom that a steady RDM may have.
 In fact, as is well known, a generic complex $2\times 2$ matrix has $8$ free real parameters.
 The unit trace and hermiticity of the RDM reduce the number of free real parameters to $3$.
 Thus, when the relation Eq.(\ref{main-result}) is valid,
 the steady RDM has approximately only one free real parameter.

 Thirdly, the relation given in Eq.(\ref{main-result}) may be useful in the study of decoherence.
 At least it suggests that, under dissipative interactions,  the following naive picture may need some modification,
 i.e., decoherence may always reduce offdiagonal elements of the RDM in the energy basis to small values,
 as long as the environment is sufficiently large and undergoes a sufficiently irregular motion.
 In fact, under a dissipative interaction,
 the total time evolution continuously brings ``small pieces'' of the branch $|\E_\beta(t)\rangle$
 into another branch $|\E_\alpha(t)\rangle$, and vice versa.
 This may generate a nonnegligible offdiagonal element.

 In particular, Eq.(\ref{main-result}) may be useful in the study of the important concept of
 preferred (pointer) basis, 
 on which the RDM becomes approximately diagonal at long times in certain robust way
 \cite{zurek2003decoherence,Zurek-ps,pra08-ps,WHG12,pre14-ps,Raedt08}.
 Since a RDM can always be diagonalized, robustness plays a key role in this concept.
 A merit of Eq.(\ref{main-result}) is that it does not depend on many detailed properties of the interaction
 and of the initial state of the environment; in other words, it possesses some robust feature.
 In future investigation, it should be of interest to study relevance of this type of robustness
 to the concept of preferred basis.

 Fourthly, one may consider a generic form of the interaction Hamiltonian $H^I$, which is written as
\begin{equation}\label{HI-g}
H^I = \sum_\nu H^{IS,\nu}\otimes H^{I\E,\nu}.
\end{equation}
 It is not difficult to verify that now Eq.(\ref{tls-rho}) takes the following form,
\begin{gather}\label{tls-rho-many}
\overline{\rho}_{\alpha\beta}^S  = \sum_\nu \eta_d^\nu \ov{F}^\nu_{\beta \alpha}
 +\eta_r^\nu  (\ov{F}^\nu_{\beta\beta}- \ov{F}^\nu_{\alpha\alpha})
 \ \  (\alpha \ne \beta),
\end{gather}
 where
\begin{gather} \label{HIE-ab-nu}
F^\nu_{\alpha\beta}(t) = \langle \E_\alpha(t)|H^{I\E,\nu}|\E_\beta(t)\rangle.
\end{gather}
 In principle, discussions given in Sec.~\ref{sect-main}
 can be generalized to this generic case, giving similar, while, more complex relations.

 Moreover,  it would be of interest to study the possibility of applying the
 approach used in this paper to central systems with $d_S>2$.
 This generalization is in principle possible, but,
 it seems much more complicated than the above-discussed generalization
 of the interaction Hamiltonian for a TLS.
 A key problem is to find conditions under which $\ov F_{\alpha\beta}$
 may be directly related to elements of $\ov\rho^S$.

\acknowledgements

 The authors are grateful to J. Gong and G. Benenti for valuable suggestions.
 This work was partially supported by the Natural Science Foundation of China under Grant
 Nos.~11275179, 11535011, and 11775210.



\begin{thebibliography}{99}
\bibitem{foot-puredephas} This is different from that reported in
Refs.\cite{addis2014coherence,roszak2015decoherence,zhang2015role}, in which the environments
are not quantum chaotic systems.
\end{thebibliography}
\end{document}